\documentclass[acmsmall,screen]{acmart}

\usepackage{natbib}

\usepackage{amsmath}
\usepackage{graphicx}
\usepackage{textcomp}
\usepackage{color}
\usepackage{xcolor}
\usepackage{makecell}
\usepackage{enumitem}
\usepackage{wrapfig}

\usepackage{amssymb}

\usepackage{amsfonts}

\usepackage{xspace}
\usepackage{multirow}
\usepackage{framed}
\usepackage{tcolorbox}
\usepackage{diagbox}
\usepackage{pifont}
\usepackage{subcaption}

\usepackage{colortbl}
\usepackage{url} 
\usepackage{listings}

\usepackage{wrapfig}

\usepackage{algorithmic}
\usepackage[linesnumbered,ruled,vlined]{algorithm2e}

\usepackage{tikz}

\usepackage{float}
\newfloat{codelist}{t}{lop}

\AtBeginDocument{%
  }

\setcopyright{acmlicensed}
\copyrightyear{2026}
\acmYear{2026}
\acmDOI{XXXXXXX.XXXXXXX}

\acmConference[FSE'26]{The ACM International Conference on the Foundations of Software Engineering (FSE)}{July 5--9, 2026}{Montreal, Canada}

\acmISBN{978-1-4503-XXXX-X/2018/06}

\newcommand{\code}[1]{\texttt{#1}}

\newcommand{\ourtoolraw}{SMART}
\newcommand{\ourtool}{\ourtoolraw{}\xspace}

\newcommand{\tabref}[1]{Table~\ref{tab:#1}}
\newcommand{\tablabel}[1]{\label{tab:#1}}
\newcommand{\figref}[1]{Figure~\ref{fig:#1}}
\newcommand{\figlabel}[1]{\label{fig:#1}}

\newcommand{\smalltitlecolon}[1]{{\noindent \bf  {#1}:\ }}

\newcounter{num}

\lstdefinestyle{Java}{ 
	language=Java,
	basicstyle=\scriptsize\ttfamily, 
	breakatwhitespace=false, 
	breaklines=true, 
	captionpos=b, 
	commentstyle=\color[rgb]{0.0, 0.5, 0.69},
	deletekeywords={}, 
	escapeinside={<@}{@>},
	firstnumber=1, 
	frame=lines, 
	frameround=tttt, 
	keywordstyle={[1]\color{blue!90!black}},
	keywordstyle={[3]\color{red!80!orange}},
	morekeywords={String,int}, 
	numbers=none, 
	numbersep=-8pt, 
	numberstyle=\tiny\color[rgb]{0.1,0.1,0.1}, 
	rulecolor=\color{black}, 
	showstringspaces=false, 
	showtabs=false, 
	stepnumber=1, 
	stringstyle=\color[rgb]{0.58,0,0.82},
	tabsize=2, 
	backgroundcolor=\color{white}
}

\begin{document}

\title{Boosting LLMs for Mutation Generation}

\author{Bo Wang}
\orcid{0000-0001-7944-9182}
\affiliation{
  \institution{Beijing Jiaotong University}
  \city{Beijing}
  \country{China}
}
\affiliation{
  \institution{Beijing Key Laboratory of Traffic Data Mining and Embodied Intelligence}
  \city{Beijing}
  \country{China}
}
\email{wangbo\_cs@bjtu.edu.cn}

\author{Ming Deng}
\orcid{0009-0007-7193-0087}
\affiliation{
  \institution{Beijing Jiaotong University}
  \city{Beijing}
  \country{China}
}
\email{24120317@bjtu.edu.cn}

\author{Mingda Chen}
\orcid{0009-0002-6283-1291}
\affiliation{
  \institution{Beijing Jiaotong University}
  \city{Beijing}
  \country{China}
}
\email{23120337@bjtu.edu.cn}

\author{Chengran Yang}
\authornote{Corresponding author.}
\orcid{0000-0001-6100-8127}
\affiliation{
  \institution{Singapore Management University}
  \country{Singapore}
}
\email{cryang@smu.edu.sg}

\author{Youfang Lin}
\orcid{0000-0002-5143-3645}
\affiliation{
  \institution{Beijing Jiaotong University}
  \city{Beijing}
  \country{China}
}
\affiliation{
  \institution{Beijing Key Laboratory of Traffic Data Mining and Embodied Intelligence}
  \city{Beijing}
  \country{China}
}
\email{yflin@bjtu.edu.cn}

\author{Mark Harman}
\orcid{0000-0002-5864-4488}
\affiliation{
  \institution{University College London}
  \city{London}
  \country{United Kingdom}
}
\email{mark.harman@ucl.ac.uk}

\author{Mike Papadakis}
\orcid{0000-0003-1852-2547}
\affiliation{
  \institution{University of Luxembourg}
  \city{Luxembourg}
  \country{Luxembourg}
}
\email{michail.papadakis@uni.lu}

\author{Jie M. Zhang}
\orcid{0000-0003-0481-7264}
\affiliation{
  \institution{King's College London}
  \city{London}
  \country{United Kingdom}
}
\email{jie.zhang@kcl.ac.uk}

\renewcommand{\shortauthors}{Wang et al.}

\begin{abstract}
LLM-based mutation testing is a promising testing technology, but existing approaches typically rely on a fixed set of mutations as few-shot examples or none at all.
This can result in generic low-quality mutations,  missed context-specific mutation patterns, substantial numbers of redundant and uncompilable mutants, and limited semantic similarity to real bugs.
To overcome these limitations, we introduce  \ourtool (Semantic Mutation with Adaptive Retrieval and Tuning).
\ourtool integrates retrieval-augmented generation (RAG) on a vectorized dataset of real-world bugs, focused code chunking, and supervised fine-tuning using mutations coupled with real-world bugs.
We conducted an extensive empirical study of \ourtool using 1,991 real-world Java bugs from the Defects4J and ConDefects datasets, comparing \ourtool to the state-of-the-art LLM-based approaches, LLMut and LLMorpheus.

The results reveal that \ourtool substantially improves mutation validity, effectiveness, and efficiency (even enabling small-scale 7B-scale models to match or even surpass large models like GPT-4o).
We also demonstrate that \ourtool significantly improves downstream software engineering applications, including test case prioritization and fault localization.
More specifically, \ourtool  
improves validity (weighted average generation rate) from 42.89\% to 65.6\%.
It raises the non-duplicate rate from 87.38\% to 95.62\%, 
and the compilable rate from 88.85\% to 90.21\%.
In terms of effectiveness, it achieves a real bug detection rate of 92.61\% (vs.\ 57.86\% for LLMut) and improves the average Ochiai coefficient from 25.61\% to 38.44\%.
For fault localization, \ourtool ranks 64 more bugs as Top-1 under MUSE and 57 more under Metallaxis.

\end{abstract}




\maketitle

\section{Introduction} \label{sec:intro}

Mutation testing~\cite{hamlet1977testing,demillo1978hints} is widely regarded as one of the most fundamental dynamic techniques in software testing and program analysis. 
By systematically introducing artificial faults, known as \textit{mutations}, into the program under test, mutation testing enables researchers and practitioners to assess the adequacy of test suites~\cite{jia2010analysis}, prioritize effective test suites~\cite{do2006use,lou2015mutation,shin2019empirical,chen2025camus}, or localize buggy statements/methods~\cite{moon2014ask,papadakis2015metallaxis,wang2025systematic}.
In recent years, mutation testing has gained widespread adoption in the industry. 
Google and Meta have reported their successful experience in deploying mutation testing~\cite{harman2025mutation,petrovic2021practical,beller2021would}.
The effectiveness of mutation testing, however, fundamentally depends on the quality of the generated mutations. Generating effective and diverse mutants is thus critical not only for evaluating test suites, but also for supporting downstream applications such as test case prioritization and fault localization.

Recently, the rapid advancement of Large Language Models (LLMs) has opened new opportunities for mutation testing~\cite{wang2024software,chen2024deep}.
Due to their strong code understanding and generation capabilities, LLMs have been explored to automatically generate more realistic and semantically meaningful mutants. 
Both academic research~\cite{degiovanni2022mubert,ibrahimzada2023automated,tip2025llmorpheus,wang2024exploratory} and industrial practice~\cite{harman2025mutation} have begun to investigate LLM-based mutation generation as a promising alternative to traditional approaches.
Despite encouraging progress, existing LLM-based approaches still suffer from several major limitations.
First, they rely on either a fixed set of few-shot examples as in-context examples or no examples at all, making it difficult for models to adapt to diverse program contexts.
For example, BugFarm~\cite{ibrahimzada2023automated} and LLMorpheus~\cite{tip2025llmorpheus} rely solely on prompts, whereas LLMut~\cite{wang2024exploratory} incorporates a fixed set of real-world bugs as few-shot examples to instruct LLMs.
Second, they typically restrict the code context to the surrounding focal method, which blinds the model to crucial program-wide dependencies.
All existing LLM-based mutation generation approaches directly adopt the whole focal method as context.
Due to these limitations, the generated mutations still contain a substantial proportion of redundant and uncompilable cases, with limited semantic similarity to real bugs~\cite{wang2024exploratory}.

To address these limitations, we present \textbf{\ourtool}  (\textbf{S}emantic \textbf{M}utation with \textbf{A}daptive \textbf{R}etrieval and \textbf{T}uning), a novel LLM-based mutation generation approach that integrates retrieval-augmented generation (RAG)~\cite{lewis2020retrieval}, code chunking, and fine-tuning~\cite{chen2021evaluating}.
Our approach is designed to enhance both the relevance and effectiveness of generated mutations. 
First, to select context-aware few-shot examples, we build a vectorized dataset of real-world bugs and employ RAG to retrieve the most relevant ``bug–fix'' pairs for the target method.
We collect 130,000 real-world single-hunk bugs with corresponding patches, and embed the patched versions as the keys for retrieval.
This ensures that the LLM benefits from in-context examples that are semantically aligned with the mutation task at hand. 
Second, rather than mutating the entire method in a monolithic way, we partition the focal method into function-coherent code chunks and generate mutations at the chunk level, which improves focus and diversity.
Finally, to adapt LLMs more closely to the task of producing semantically effective mutations, we apply supervised fine-tuning (SFT) using effective mutations that are coupled with real-world bugs.
The training set comprises 13,760 mutations that are coupled with real bugs from the prior study~\cite{wang2024exploratory}.
To prevent data leakage, all the training mutations are from projects beyond our evaluation.
By combining retrieval, structural code decomposition, and model adaptation, \ourtool aims to advance the state of LLM-based mutation generation significantly.

To assess the quality of the mutations generated by \ourtool, we conducted an extensive empirical study on 1,991 real-world Java bugs, including 701 from Defects4J~\cite{just2014defects4j} and 1,290 from ConDefects~\cite{wu2023condefects}.
We evaluated \ourtool using five open-source LLMs (DeepSeek-Coder-6.7B, Llama-3.1-8B, Qwen-2.5-7B, Qwen-2.5-14B, and Qwen-2.5-32B), and additionally tested GPT-4o with RAG and code chunking enabled.
For comparison, we included two state-of-the-art LLM-based mutation generation approaches: LLMorpheus~\cite{tip2025llmorpheus} and LLMut~\cite{wang2024exploratory}.
Our evaluation covers a wide range of dimensions. We first examine the effectiveness of generated mutations, i.e., how closely they couple with real-world bugs. We then assess their validity, since LLMs inevitably produce redundant or incompilable mutants.
Finally, we evaluate their impact on downstream applications, including mutation-based test case prioritization and fault localization.

The results show that \ourtool substantially improves both the \textit{validity} and \textit{effectiveness} of LLM-generated mutations.
For validity, \ourtool raises the weighted average generation rate from 42.89\% (LLMut) to 65.6\% ($\uparrow 52.95\%$; +22.71 PPs), increases the non-duplicate rate from 87.38\% (LLMut) and 85.87\% (LLMorpheus) to 95.62\% ($\uparrow 9.42\%$--11.36\%), and improves the compilable rate from 88.85\% (LLMut) and 78.43\% (LLMorpheus) to 90.21\% ($\uparrow 1.53\%$--15.0\%).
For effectiveness, \ourtool achieves a weighted average real bug detection rate of 92.61\%, compared to 57.86\% for LLMut and 31.99\% for LLMorpheus, and raises the average Ochiai coefficient to 38.44\% ($\uparrow 50.1\%$ and 277.6\% over baselines).
On downstream applications, \ourtool enhances mutation-based fault localization by ranking more bugs in Top-$k$ and reducing MAR/MFR (e.g., 143 Top-1 bugs by MUSE vs. 106 for LLMut and 11 for LLMorpheus).
Notably, our approach enables 7B-scale models to achieve performance competitive with GPT-4o, making it suitable for scenarios with limited computational resources.

In the discussion section, we further analyze the computational cost and robustness of \ourtool. Specifically, we report token consumption during mutation generation, as well as the overhead introduced by RAG and fine-tuning. Building the RAG index requires approximately 10 minutes as a one-time offline cost, while retrieval incurs an average overhead of 6.5 seconds per bug when chunking is enabled.
We also examine the impact of different similarity metrics used in RAG and compare cosine similarity, dot-product similarity, and Euclidean distance. Our results show that Euclidean distance consistently achieves the best performance on semantic-related metrics and downstream tasks. These analyses provide a more comprehensive understanding of its practical cost and sensitivity to configuration choices.


In summary, our paper makes the following contributions:
\begin{itemize}[leftmargin=*]
    \item We propose a novel LLM-based mutation generation approach, \ourtool, which incorporates retrieval-augmented generation (RAG), code chunking, and fine-tuning.
    
    \item We conduct an extensive study to evaluate \ourtool, measuring its mutation validity, effectiveness, and applications in mutation-based test case prioritization and fault localization. The results demonstrate the significant superiority of \ourtool compared with the two state-of-the-art LLM-based approaches.
    
    \item Our implementation, the checkpoints of fine-tuned models, and high-quality Java mutations are publicly available.
\end{itemize}
\section{Background} \label{sec:back}
In this section, we introduce the necessary concepts of mutation testing, including metrics for evaluating the quality of mutations and their downstream applications.

\subsection{Mutation Testing}
Mutation testing is a fault-injection technique that assesses test quality by injecting small syntactic changes, called \emph{mutations}, into the program~\cite{hamlet1977testing,demillo1978hints,jia2010analysis,papadakis2019mutation}. 
A mutant is \emph{killed} if test outputs differ from the original program; otherwise it \emph{survives}, exposing weaknesses in the test suite. 
The \emph{mutation score}, i.e., the proportion of killed mutants, is a standard measure of test effectiveness.

Traditional mutation testing relies on rule-based operators (e.g., replacing arithmetic or relational operators)~\cite{just2011major,coles2016pit}, which, while effective, often fail to capture the diversity of real faults.  
To overcome this, learning-based approaches such as DeepMutation~\cite{tufano2019learning} and LEAM~\cite{tian2022learning,tianleam++} leverage deep learning to guide mutation generation.  
More recently, LLMs have emerged as a new direction, offering the potential to generate semantically richer and more realistic mutants.  
BugFarm~\cite{ibrahimzada2023automated} leverages LLMs to synthesize hard-to-kill artificial bugs for test evaluation.  
LLMorpheus~\cite{tip2025llmorpheus} explores the feasibility of using LLMs for mutation generation.  
LLMut~\cite{wang2024exploratory} systematically studies the validity and effectiveness of LLM-generated mutations in Java.

\subsection{Evaluating Mutations} \label{sec:evaluating-mutations}
The usefulness of mutation testing largely depends on the quality of the generated mutants.
To better characterize this quality, we first categorize the mutations generated by LLMs~\cite{wang2024exploratory}. 
Let all LLM-generated mutations be denoted as the set $A$, which may include invalid mutations. 
We denote the subset of compilable mutations as $C$. 
Within $C$, two subsets of mutations are useless: duplicate mutations ($D$), which are syntactically identical to the original program or to other mutants, and equivalent mutations ($E$), which are syntactically different but semantically indistinguishable from the original program.
Formally, $(C \supseteq D) \wedge (C \supseteq E) \wedge (D \cap E = \phi)$. 
Therefore, the ideal set of useful mutants for mutation testing is $(C - D - E)$, excluding both duplicates and equivalents that do not contribute to exposing weaknesses in the test suite.
However, because determining whether a mutant is equivalent is undecidable~\cite{budd1982two}, and the proportion of equivalent mutants is typically negligible in practice ($|E| \ll |C - D|$)~\cite{wang2024exploratory,tian2024large,tip2025llmorpheus}, we approximate the ideal set of useful mutants by $(C - D)$ for downstream mutation testing applications.
For instance, the mutation score $MS$ is calculated as follows:
\begin{equation}
    MS = \frac{| K | }{ | C - D | } \approx \frac{| K | }{ | C - D - E |} \quad  (|E| \ll |C - D|),
\end{equation}
where $K$ is the set of killed mutations.

\subsubsection{Validity}
Invalid mutants (e.g., uncompilable code or duplicates of the original program) provide little value and may bias evaluation results.
Therefore, validity is commonly measured using two indicators:  
(1) the \textit{compilability rate}, i.e., the proportion of mutants that can be successfully compiled (i.e., $CR =  {|C|} / {|A|}$);
(2) the \textit{duplication rate}, i.e., the proportion of mutants that are syntactically identical to the original program or other existing mutants (i.e., $DR = {|D|} / {|A|}$).
A high-quality mutation generation approach should maximize compilability while minimizing duplication and equivalence.

\subsubsection{Effectiveness}
The effectiveness of mutations reflects how well the generated mutants approximate real-world faults. 
Evaluating effectiveness is crucial, since only mutants that resemble real bugs can provide meaningful insights into the fault-detection capability of a test suite and the practical value of mutation testing. 
Particularly, for LLM-generated mutations, effectiveness involves examining whether the introduced changes capture fault characteristics that are commonly observed in real defects. 

Specifically, the effectiveness of mutations comprises three aspects.
First, the \textbf{real-bug detection capability}, which evaluates whether bug-triggering test cases that reveal real faults are also able to detect the generated mutants~\cite{just2014mutants}.
Second, the \textbf{fault coupling rate}, i.e., the proportion of generated mutants that couple with real bugs.
Third, the \textbf{semantic similarity} between a mutant and its corresponding real bug, which can be quantified using the \textbf{Ochiai coefficient}~\cite{ojdanic2023syntactic,wang2024exploratory}.

Formally, let $m$ and $b$ be a mutant and its corresponding real bug, and let $fT_m$ and $fT_b$ be the sets of failing test cases for $m$ and $b$, respectively. 
The Ochiai coefficient between $m$ and $b$ is defined as:
\begin{equation}
    Ochiai(m,b) = \frac{| fT_m \cap fT_b | }{ \sqrt{| fT_m | \times |fT_b|} }.
\end{equation}
A coefficient closer to 1 indicates that the mutant is more semantically similar to the real bug.
For a real bug $b$, its Ochiai coefficient is calculated as the mean coefficient across all de-duplicated mutants $(C_b - D_b)$ generated for it:
\begin{equation}
    Ochiai_b = \frac{ \sum_{m \in (C_b - D_b)} Ochiai(m,b) }{ | C_b - D_b | },
\end{equation}
where $C_b$ and $D_b$ are the corresponding sets $C$ and $D$ of the bug $b$, respectively.
Given a mutation generation approach and a set of bugs $B$, the overall effectiveness is quantified by the Average Ochiai Coefficient (AOC):
\begin{equation}
    AOC = \frac{\sum_{b \in B} {Ochiai_b} }{|B|}.
    \label{eq:aoc}
\end{equation}

\subsection{Downstream Applications of Mutation Testing}
In this paper, we consider the two widely studied applications of mutation testing, \textit{mutation-based test case prioritization} and \textit{mutation-based fault localization}.

\subsubsection{Mutation-Based Test Case Prioritization} \label{sec:mbtcp}
Test case prioritization (TCP) aims to schedule the execution order of test cases so that faults can be detected as early as possible, and has been extensively studied in the literature~\cite{yoo2012regression,rothermel2001prioritizing}.

Mutation-based TCP~\cite{do2006use,lou2015mutation,shin2019empirical}, which leverages information from the kill matrix, has shown its effectiveness compared with coverage-based methods.
Recently, the TCP approaches, GRK, GRD, and HYB-$\omega$, have been widely studied~\cite{do2006use,tian2022learning,tianleam++}.

\textbf{GRK} (Greedy Kill) iteratively selects the test case that kills the maximum number of additional mutants, aiming to separate faulty behavior from the original program as early as possible.  
\textbf{GRD} (Greedy Distinguish) iteratively selects the test case that distinguishes the maximum number of additional mutants, where two mutants are distinguished if their execution produces different outputs. Thus, GRD attempts to maximize the diversity of fault-revealing behaviors.  
\textbf{HYB-$\omega$} combines both strategies by selecting, at each step, the test case that maximizes a weighted sum of additionally killed and additionally distinguished mutants. When $\omega=1$, HYB-$\omega$ degenerates to GRK; when $\omega=0$, it degenerates to GRD.

\subsubsection{Mutation-Based Fault Localization} \label{sec:mbfl}
Mutation-based fault localization (MBFL) techniques leverage the behavior of mutants to estimate the suspiciousness of program entities. 
The intuition is that if the execution of a mutant impacts failing test cases more often than passing ones, the associated statement is likely to be faulty.
Among various MBFL methods, \textsc{MUSE}~\cite{moon2014ask} and \textsc{Metallaxis}~\cite{papadakis2015metallaxis} are two widely studied techniques.

\smalltitlecolon{MUSE} 
MUSE first assigns a suspiciousness score to each mutant $m$ using both failing and passing test cases, shown as follows: 
\begin{equation}
   S(m) = failed(m) - \frac{f2p}{p2f} \cdot passed(m) ,
\end{equation}
where $failed(m)$ and $passed(m)$ denote the number of test cases that fail or pass on the original program but yield the opposite outcome on mutant $m$. 
$f2p/p2f$ represents the number of test cases that flip from failing to passing or from passing to failing due to any mutant. 
The suspiciousness of a statement $s$ is then calculated as the average suspiciousness of all mutants located at $s$.

\smalltitlecolon{Metallaxis} 
In contrast, Metallaxis computes suspiciousness by normalizing with respect to the total number of failing test cases, shown as follows:
\begin{equation}
    S(m) = \frac{failed(m)}{\sqrt{totalfailed \cdot (failed(m)+passed(m))}}
\end{equation}
Here, $totalfailed$ is the number of failing test cases on the original program, while $failed(m)$ and $passed(m)$ are defined similarly as in MUSE. 
The suspiciousness of a statement $s$ is determined as the maximum suspiciousness score among all mutants associated with $s$.



\section{Methodology}
\label{sec:methodology}

In this section, we introduce \ourtool, a novel LLM-based mutation generation approach that integrates logic-based code chunking (Section~\ref{sec:chunking}), RAG-driven few-shot example retrieval (Section~\ref{sec:rag}), prompt engineering for the specified task (Section~\ref{sec:prompt}), and supervised fine-tuning (Section~\ref{sec:ft}).

\subsection{Overview}

\begin{figure}[t]
    \centering
    \includegraphics[width=0.85\linewidth]{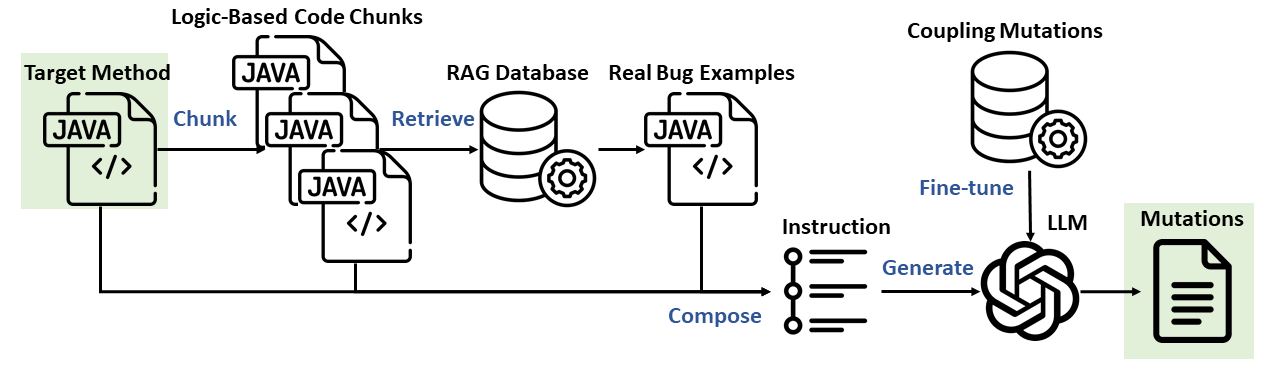}
    \caption{The Overview of Mutation Generation Process of \ourtool}
    \figlabel{overview}
\end{figure}

\figref{overview} illustrates the overall workflow of \ourtool, which integrates three key techniques (i.e., logic-based code chunking, retrieval-augmented generation (RAG), and supervised fine-tuning (SFT)), into a unified pipeline.

Given a focal Java method as input, we first apply logic-based code chunking to partition the focal method into semantically coherent segments.
This decomposition reduces the complexity of mutation generation and allows the model to concentrate on finer-grained program logic. 
Next, to provide context-aware few-shot examples, we build a vectorized repository of real-world bug–fix pairs and employ RAG to retrieve the most relevant examples for the target method. 
These retrieved examples are then assembled into the prompt to guide LLMs in generating more effective mutations.
In parallel, we further adapt LLMs to the mutation generation task through supervised fine-tuning on mutations that are semantically coupled with real-world bugs.
Finally, the fine-tuned model generates mutations at the chunk level, and the results are aggregated to form the candidate mutation set for further evaluation.

\begin{figure}[t]
    \centering
    \includegraphics[width=\linewidth]{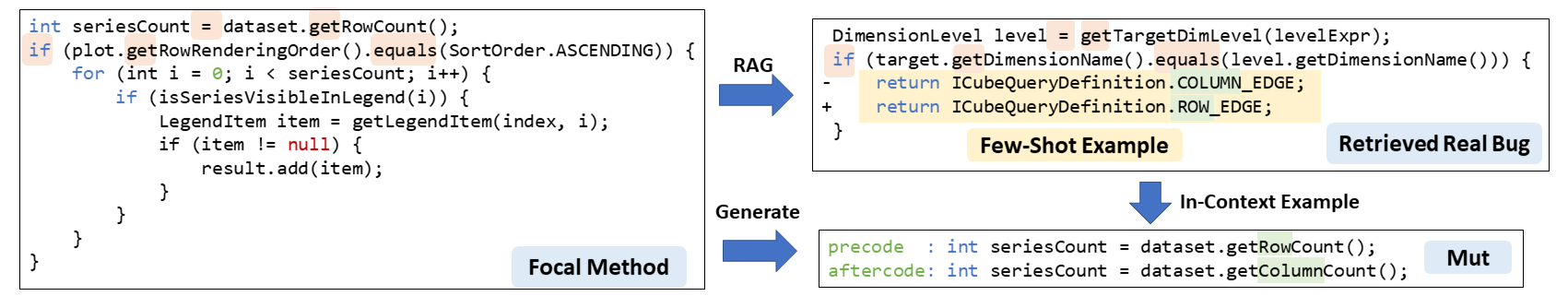}
    \caption{The Example Mutation of \ourtool}
    \figlabel{example}
\end{figure}

\figref{example} illustrates a mutation example of \ourtool.
Given a code chunk from the focal method (left), our RAG component retrieves a real bug (right) whose code is similar to the chunk: both contain (1) an assignment to a local variable, and (2) an \code{if}-condition composed of chained \code{get*()} and \code{equals()} calls. 
In the retrieved bug, the fix swaps the semantic roles of \code{COLUMN} and \code{ROW} (\code{COLUMN\_EDGE} $\rightarrow$ \code{ROW\_EDGE}). 
We feed this instance as a few-shot example to the LLM, which then produces a semantically analogous mutation for the target chunk, which changes 
\code{int seriesCount = dataset.getRowCount();} to
\code{int seriesCount = dataset.getColumnCount();}.
This mutation captures a realistic \emph{row/column} semantic swap rather than a superficial token edit, thereby strengthening test suites by exposing tests that implicitly assume a specific orientation (row vs.\ column). 
The example highlights the capability of \ourtool in generating semantically effective mutations.

\subsection{Logic-Based Code Chunking} \label{sec:chunking}
Code chunking has proven effective in LLM-related tasks~\cite{wang2023codet5,wang2024hits}, as it reduces the input size and allows LLMs to focus more accurately on the relevant code context.
However, naively splitting a focal method by a fixed chunk size may break the semantic structure of the code and lead to incoherent inputs.
To address this issue, we design a \emph{logic-based code chunking} that partitions methods along their natural control-flow and declaration boundaries.
The procedure is illustrated in Algorithm~\ref{alg:chunking}.

The algorithm takes a focal method $M$ as input and outputs a set of chunks $C$.
We first extract the line numbers of all statements in $M$ and parse the method into its abstract syntax tree (AST).
Next, we collect the target nodes that represent control-flow structures (\texttt{if}, \texttt{for}, \texttt{while}, \texttt{do-while}, and \texttt{try-catch}).
These nodes are sorted in descending order of their line numbers to ensure bottom-up processing of nested statements.

For each target node $t$, we identify the corresponding code lines $l_t$.
If these lines have not yet been collected, we create a chunk containing the code of $t$ and extend it with the declaration statements that immediately precede $t$ to preserve semantic completeness.
We then mark these lines as processed by removing $l_t$, and the declaration lines $l_d$ from the set $\mathcal{L}$.

After handling all control-flow nodes, the remaining lines in $\mathcal{L}$ are grouped into consecutive segments.
Each segment forms a separate chunk, ensuring that every line of the focal method is included in exactly one chunk.
Finally, the algorithm returns the set of all chunks $C$, which collectively preserve the logical structure of the original method while reducing the complexity of mutation generation.

\begin{algorithm}[tp]
\small
\caption{The Chunking Algorithm of \ourtool.}
\label{alg:chunking}

\KwIn{The focal method: $M$}
\KwOut{The set of code chunks: $C$}

$C \leftarrow \phi$ \\

$\mathcal{L} \leftarrow \operatorname{get\_line\_num\_set}(M)$ \\

$\mathcal{A} \leftarrow \operatorname{parse\_ast}(M)$ \\

$\mathcal{T} \leftarrow \operatorname{get\_target\_nodes}(\mathcal{A}, \{ \text{``IfStmt"}, \text{``ForStmt"}, \text{``WhileStmt"}, \text{``DoWhileStmt"}, \text{``TryStmt"} \})$ \\

$\mathcal{T} \leftarrow \operatorname{sort\_by\_line\_descending}(\mathcal{T})$ \\

\For{$t \in \mathcal{T}$}{
    $l_t \leftarrow \operatorname{get\_line\_num\_set}(t)$ \\
    \If{$\mathcal{L} \cap l_t \neq \phi$} {
        $c \leftarrow \operatorname{get\_code\_by\_lines}(\mathcal{L} \cap l_t$) \\
        $d \leftarrow \operatorname{get\_preceding\_decl\_stmts}(t)$ \\
        $l_d \leftarrow \operatorname{get\_line\_num\_set}(d)$ \\
        $c \leftarrow c \cup \operatorname{get\_code\_by\_lines}(\mathcal{L} \cap l_d$) \\
        $C \leftarrow C \cup \{c\}$ \\
        $\mathcal{L} \leftarrow (\mathcal{L} - l_t - l_d)$ \\
    }
}
\For{$seg \in \operatorname{split\_into\_consecutive\_segments}(\mathcal{L})$}{
    $c \leftarrow \operatorname{get\_code\_by\_lines}(seg)$ \\
    $C \leftarrow C \cup \{c\}$ \\
}
\Return $C$
\end{algorithm}

\subsection{Context-Aware Few-Shot Examples Retrieval via RAG} \label{sec:rag}
To provide the LLM with relevant guidance when generating mutations, we design a RAG component that 
retrieves context-aware few-shot examples from a repository of real-world bugs.  

We collect 130,000 single-hunk Java bugs from open-source repositories. 
Each bug is represented as a pair of \textit{pre-fix} and \textit{post-fix} code, i.e., the buggy code before the patch and the corresponding fixed code after the patch. 
These bug–fix code pairs serve as the source of potential few-shot examples.
To enable efficient semantic retrieval, we embed each buggy method using the encoder of CodeT5~\cite{wang2021codet5}, a pre-trained model specialized for code representation learning. 
The embeddings are stored in a vectorized dataset, enabling scalable similarity search.

Given a code snippet (e.g., an entire method or a code chunk) to be mutated, we compute its embedding using the same encoder. 
We then perform a similarity search against the repository using \textit{Euclidean Distance} to measure the closeness between embeddings. 
The Top-$N$ most similar bug–fix pairs are selected as context-aware few-shot examples, where $N$ is a hyperparameter.

\subsection{Prompt Design} \label{sec:prompt}

\begin{figure}[t]
\begin{tcolorbox}[boxrule=1pt, left=4pt, right=4pt, top=2pt, bottom=2pt]

\textbf{[Instruction]}: Below is the original Java method, followed by a specific code chunk extracted from it. Your task is to generate \textbf{\{N\}} mutant versions by applying single-line mutations only within the code chunk.
Note: In software engineering, a mutant refers to a variant of the original program created by introducing small syntactic changes, which are typically used for mutation testing.

\textbf{[Entire Focal Method]}: \textbf{\{JAVA METHOD\}}

\textbf{[The Current Chunk]}: Only mutate these lines: \textbf{\{Code Chunk\}}

\textbf{[Few-Shot Examples]}: <json> \textbf{\{Examples\}} </json>

\textbf{[Output Instructions]}: 

1. A mutation can only occur on one line. 

2. Your output must be like: <json> [ \{ ``precode": ``", ``aftercode": ``" \} ] </json>. The "precode" represents the line of code before mutation, and it can't be empty, "aftercode" represents the line of code after mutation. Note that you may need to generate multiple pairs of "precode" and "aftercode". 

3. Prohibit generating mutants that are identical to the original code (precode) or duplicate any previously generated mutants. 

4. Output all mutations in JSON format, ensuring they are wrapped in <json></json> tags.

\end{tcolorbox}
\caption{The Prompt Template of \ourtool for Mutation Generation}
\figlabel{prompt}
\end{figure}

To guide LLMs in generating high-quality mutations, we design a structured prompt template, as illustrated in \figref{prompt}. 
The prompt consists of four main components: \emph{Instruction}, \emph{Focal Method}, \emph{Code Chunk}, and \emph{Few-Shot Examples}, together with explicit \emph{Output Instructions}.

We first define the task by natural language instruction.
To provide the global context, we include the entire focal method. 
This ensures that the LLM can understand the broader semantics of the code and avoid generating inconsistent or irrelevant changes.
Then, we highlight the specific code chunk to be mutated. By restricting modifications to this chunk, we reduce ambiguity in the mutation task and force the LLM to focus on a semantically coherent portion of the method.
To further enhance context-awareness, we integrate bug–fix pairs retrieved by RAG as few-shot examples.
These examples are serialized in JSON format to maintain consistency and to make it easier for the LLM to learn the structure of valid input–output mappings.
Finally, we explicitly constrain the output format. 
The LLM must generate mutants where each mutation alters only one line, and the output is required to follow a JSON-based structure containing the \texttt{precode} and \texttt{aftercode} fields. 
This design facilitates automatic parsing and integration of generated mutants into downstream mutation-testing pipelines.

\subsection{Fine-Tuning} \label{sec:ft}
In addition to the RAG component, which provides relevant, context-aware examples at inference time, we employ supervised fine-tuning (SFT) to specialize the model parameters for mutation generation.
The primary goal of SFT in \ourtool is to align the model more closely with producing semantically effective mutations by training it on mutations that have been empirically validated to couple with real-world bugs.

Each mutation pair, consisting of the original code (``precode'') and the corresponding mutation code (``aftercode''), serves as a golden example of a meaningful semantic change (the details of the training data will be presented in  Section~\ref{sec:training-data}).
We format these pairs into high-quality training instances that follow the same structure as our final prompt design (see Section~\ref{sec:prompt}).

For the SFT process, \ourtool uses a standard causal language modeling objective, which is mathematically formulated as minimizing the cross-entropy loss. The goal is to adjust the model's parameters, denoted by $\theta$, to maximize the conditional probability of generating a target sequence $Y$ (the ground-truth mutation) given an input prompt $X$. 
For a single training instance $(X, Y)$, the loss function $\mathcal{L}(\theta)$ is defined as:

\begin{equation}
\mathcal{L}(\theta) = - \sum_{t=1}^{n} \log P(y_t | X, y_{1:t-1}; \theta) ,
\end{equation}
where $X$ is the input prompt, which includes the instruction, the entire focal method, and the specific code chunk to mutate; $Y = (y_1, y_2, \dots, y_n)$ is the target output sequence, which is the ground-truth JSON string representing the mutation (e.g., \texttt{\{"precode": "...", "aftercode": "..."\}}); $y_t$ denotes the $t$-th token in the target sequence $Y$; and $P(y_t | X, y_{1:t-1}; \theta)$ is the probability assigned by the model with parameters $\theta$ to the correct token $y_t$ at position $t$, given the input prompt $X$ and all preceding ground-truth tokens $y_{1:t-1}$.
By minimizing this loss over the entire training dataset, the model's parameters $\theta$ are updated via backpropagation to make the ground-truth mutations more probable, effectively aligning the model with the task of generating realistic faults. 


\section{Evaluation Design} \label{design}
In this section, we elaborate on the details of our evaluation design.

\subsection{Research Questions}

We aim to evaluate \ourtool by answering the following research questions.
\begin{itemize}[leftmargin=*]

    \item \textbf{RQ1 (Validity) Does \ourtool generate more valid mutations than existing approaches?}

    \item \textbf{RQ2 (Effectiveness) Do mutations generated by \ourtool more closely resemble real bugs than those from existing approaches?}  
    
    \item \textbf{RQ3 (Mutation-Based Test Case Prioritization) How does the use of \ourtool affect the performance of mutation-based test case prioritization compared with existing approaches?} 

    \item \textbf{RQ4 (Mutation-Based Fault Localization) How does \ourtool affect the performance of mutation-based fault localization techniques compared with existing approaches?}  

    \item \textbf{RQ5 (Ablation Study on Components)  Are the individual components useful for improving mutation generation?} 


\end{itemize}

\subsection{Studied Mutation Generation Approaches}

In this study, we compare \ourtool with two representative state-of-the-art LLM-based mutation generation approaches: 
LLMorpheus~\cite{tip2025llmorpheus} and LLMut~\cite{wang2024exploratory}.  
LLMorpheus represents one of the most recent frameworks for leveraging LLMs in mutation testing, while LLMut conducts a systematic study of mutation effectiveness and validity.  
Together, these two approaches serve as strong baselines for evaluating the performance of \ourtool.

Beyond these baselines, we further conduct an ablation study to isolate the contributions of individual components in our design.  
Specifically, we explore the following variants: 
\begin{itemize}[leftmargin=*]
    \item \textbf{\ourtoolraw$_{R}$}, which augments the vanilla LLM with RAG to search in-context few-shot example;  
    \item \textbf{\ourtoolraw$_{C}$}, which applies logic-based code chunking without retrieval or fine-tuning; 
    \item \textbf{\ourtoolraw$_{RC}$}, which combines retrieval with chunk-level mutation generation;  
    \item \textbf{\ourtoolraw$_{CF}$}, which applies supervised fine-tuning while using chunking; and  
    \item \textbf{\ourtool}, our complete approach that integrates RAG, chunking, and fine-tuning.  
\end{itemize}

For mutation generation, we evaluate a diverse set of LLMs.  
Specifically, we employ five open-source models: \textbf{DeepSeek-Coder-6.7B}~\cite{zhu2024deepseek}, \textbf{Llama-3.1-8B}~\cite{grattafiori2024llama}, \textbf{Qwen-2.5-7B}, \textbf{Qwen-2.5-14B}, and \textbf{Qwen-2.5-32B}~\cite{yang2025qwen3}.  
The three Qwen-2.5 variants share the same architecture but differ in parameter scales, which allows us to study the effect of model size under a controlled setting.
Additionally, for experiments that do not involve fine-tuning, we also include \textbf{GPT-4o}~\cite{achiam2023gpt} (i.e.,\ourtoolraw$_{R}$, \ourtoolraw$_{C}$, \ourtoolraw$_{RC}$, LLMut, and LLMorpheus).  
Note that for fairness and consistency, we adopt GPT-4o as the backend model for LLMorpheus~\cite{tip2025llmorpheus} and LLMut~\cite{wang2024exploratory}, since GPT-4o is the most widely studied model in existing LLM-based mutation testing research~\cite{tip2025llmorpheus,wang2024exploratory}.

Through these comparisons, we aim to examine not only whether \ourtool outperforms existing methods, but also how each of its components contributes to improving the quality of generated mutants and their downstream utility in mutation testing.

\subsection{Training Data and Subjects Under Test} \label{sec:training-data}
We conduct our study on two widely used benchmarks of real-world Java bugs: \textbf{Defects4J 2.0}~\cite{just2014defects4j} and \textbf{ConDefects}~\cite{wu2023condefects}. 
Defects4J 2.0 is a widely used benchmark in mutation testing research~\cite{ojdanic2023syntactic,just2014mutants,kaufman2022prioritizing,degiovanni2022mubert,tian2022learning,kim2022predictive,wang2024exploratory,chen2025camus}.
It provides a collection of historical bugs from open-source projects across diverse domains, offering a broad and representative set of real-world software faults.
In contrast, \textsc{ConDefects} is constructed from tasks in the AtCoder programming contest\footnote{\url{https://atcoder.jp}}, and thus complements Defects4J with competitive programming–style defects.
Both datasets contain real faults together with developer-written bug-revealing test cases, which provide a reliable basis for evaluating mutation-based techniques.
\tabref{dataset} presents the statistics of the two datasets.
We observe that bugs from Defects4J exhibit greater complexity compared to those in ConDefects. Specifically, they are subjected to far more developer-written tests (250.5 on average vs. 26.1 in ConDefects) and are located in methods with longer code bodies (34.15 LOC vs. 26.82 LOC), making them more challenging for mutation generation and analysis.

\begin{table}[tp]
  \centering
  \scriptsize
  \caption{The Statistics of the Used Datasets}
    \begin{tabular}{lrrrr}
    \toprule
    \textbf{Name} & \textbf{Bug \#} & \textbf{Avg. Test \#} & \textbf{Avg. Buggy Mtd. \#} & \textbf{Avg. Mtd. LOC} \\
    \midrule
    \textbf{Defects4J} & 701 & 250.5 & 2.02 & 34.15 \\
    \textbf{ConDefects} & 1290 & 26.1 & 1  & 26.82 \\
    \bottomrule
    \end{tabular}
  \tablabel{dataset}
\end{table}

For the experiments on \textit{mutation effectiveness}, \textit{mutation validity}, and \textit{mutation-based TCP}, we generate mutants from the \textit{fixed} versions of each bug. 
This allows us to measure (1) the similarity between generated mutants and their corresponding real faults, (2) the validity of the generated mutants, and (3) whether bug-revealing test cases are prioritized earlier by mutation-based TCP techniques. 
For the experiments on \textit{MBFL}, we instead generate mutants from the \emph{buggy} versions of the programs. 
This setting enables us to evaluate the impact of mutants on MBFL techniques such as MUSE~\cite{moon2014ask} and Metallaxis~\cite{papadakis2015metallaxis}. 

To avoid potential data leakage, we randomly select two projects from Defects4J 2.0, namely \textit{Lang} and \textit{Cli}, and exclude them from our evaluation.
For training, we adopt the mutation dataset constructed by LLMut~\cite{wang2024exploratory}.
In particular, to adapt LLMs toward generating semantically meaningful mutants, we use mutations that are \emph{coupled with real-world bugs}, i.e., mutants that can be killed by at least one bug-revealing test.
Our training set consists of 13,760 such coupled mutants, which are collected from diverse sources, including LLMs (e.g., DeepSeek\cite{zhu2024deepseek}, GPT-4o~\cite{achiam2023gpt}, and CodeLlama~\cite{roziere2023code}) and mutation testing tools (e.g., LEAM~\cite{tian2022learning}, Major~\cite{just2011major}, and $\mu$BERT~\cite{degiovanni2022mubert}).

\subsection{Mutation Generation Setups}
For each real bug in the studied datasets, we generate mutants from the corresponding methods.
For RQ1–RQ3, we use fixed versions, while for RQ4, we use buggy versions.
If a bug spans multiple methods, we invoke the generation procedure independently for each method. 
Given a focal method (or its logic-based chunks, in the case of \ourtool), we configure the LLM to generate $N$ mutations, where $N$ equals the number of lines of code in the input, following common practice~\cite{wang2024exploratory}.
Following the practice of \cite{wang2024exploratory,deng2024large}, we use the \textit{Top-6} few-shot examples.
During mutation generation, all LLMs are run with their default temperature and token limitations.


\subsection{Fine-tuning Settings}
We implement \ourtool with HuggingFace Transformers\footnote{\url{https://huggingface.co/docs/transformers/en/index}} for supervised fine-tuning. 
We fine-tune all LLMs with full-parameter fine-tuning. The loss is only computed on the model responses. 
We truncate input sequences to a cutoff length of 4096 tokens to avoid out-of-memory errors.
This setting balances context retention with computational efficiency, and this threshold encompasses all training instances, ensuring maximal learning integrity.
Following standard practice, for each selected LLM, we train 3 epochs with cosine learning rate decay.
We randomly select 10\% of the dataset for warm-up. 
The base learning rate is set as $3.0\times 10^{-5}$.
Per-device batch size is set as 4 with gradient accumulation as 4.

\section{Results and Analysis} \label{sec:res}

\subsection{RQ1: Performance Comparison in the Validity of Mutations}

\subsubsection{Metrics}
As introduced in Section~\ref{sec:evaluating-mutations}, the validity of mutations reflects whether the generated code is well-formed and practically usable. 
We assess the validity of mutations generated by different LLM-based mutation approaches from three complementary perspectives:  
\begin{itemize}[leftmargin=*]
    \item \textbf{Generation Rate (Ge. R.).}  
    In both \ourtool and LLMut, the number of requested mutations is explicitly specified in the prompt (denoted as \emph{Exp.}).  
    We measure the proportion of successfully parsed mutations (denoted as \emph{Gen.}) over the expected number, i.e., $\text{Gen.} / \text{Exp.}$.  
    A higher generation rate indicates that the model more reliably follows generation instructions.  

    \item \textbf{Non-duplicate Rate (ND. R.).}  
    We compute the proportion of generated mutations that are non-redundant, i.e., not identical to the original program or to other generated mutations, which is given by $|A - D| / |A|$.
    This metric evaluates whether the model produces diverse outputs rather than trivial or repeated code.  

    \item \textbf{Compilable Rate (Com. R.).}  
    We calculate the proportion of generated mutations that can be successfully compiled, which is given by $|C| / |A|$.
    This reflects the syntactic correctness of the generated code and its practical utility in downstream mutation testing tasks.
\end{itemize}

\subsubsection{Results}

\begin{table}[tp]
  \centering
  \scriptsize
  \caption{The Validity Comparison}
    \begin{tabular}{|l|l|l|l|l|l||l|l|l|l|l|}
    \hline
    \multicolumn{1}{|c|}{\multirow{2}{*}{\textbf{Approaches}}} & \multicolumn{5}{c||}{\textbf{Defects4J (701 Bugs)}} & \multicolumn{5}{c|}{\textbf{ConDefects (1290 Bugs)}} \\
\cline{2-11}       & \textbf{Exp.} & \textbf{Gen.} & \textbf{Ge. R.} &  \textbf{ND. R.} & \textbf{Com. R.} & \textbf{Exp.} & \textbf{Ge.} & \textbf{Ge. R.} & \textbf{ND. R.} & \textbf{Com. R.} \\
    \hline
    \textbf{LLMorpheus (GPT-4o)} & -  & 4092  & - &  88.25\%  &  58.71\%  & -  & 3861  & - & 83.35\%  & 99.32\% \\
    \hline
    \textbf{LLMut (GPT-4o)} & 46873  &  6519 & 13.91\% &  95.77\%  &  83.05\%  &  33238  &  1900  & 5.72\% &  97.84\%  & \cellcolor{gray!40}99.95\% \\
    \hline
    \textbf{LLMut (DS-6.7b)} &  46873  &  3142 & 6.70\% &  55.25\%  &  68.26\%  &  33306 & 4661  & 13.99\%&  72.19\%  & 99.58\% \\
    \textbf{LLMut (LM3.1-8b)} &  46873  &  3507  & 7.48\% &  67.49\%  &  66.50\%  &  33306  & 5530 & 16.60\%  &  84.25\%  &  99.25\%  \\
    \textbf{LLMut (QW2.5-7b)} & 46873  &  7623  & 16.26\% &  83.13\%  &  70.79\%  &  33306  & 7413  & 22.26\% & 88.20\% &  99.51\% \\
    \textbf{LLMut (QW2.5-14b)} &  46873 & 10501 & 22.40\% &  86.93\%  &  75.82\%  &  33306  &  19877  & 59.68\%  & 89.69\%   &  99.73\% \\
    \textbf{LLMut (QW2.5-32b)} & 46873  & 23708 & 50.58\% &  84.96\% &  78.63\%  &  33306  &   25265  & 75.86\% &  96.37\%  & 99.70\% \\
    \hline
    \textbf{SMART$_{RC}$ (GPT-4o)} & 35979 &  17579  &  48.86\% &  \cellcolor{gray!40}97.98\%  & \cellcolor{gray!40}85.65\% &  26980 & 11680 & 43.29\%  &  \cellcolor{gray!40}99.32\%  & 99.65\% \\
    \hline
    \textbf{SMART (DS-6.7b)} & 35979 & 22012 & 61.18\% & 92.44\% & 82.84\% & 26980 & 21841 & \cellcolor{gray!40}80.95\% & 97.81\%  & 99.78\% \\
    \textbf{SMART (LM3.1-8b)} & 35979 & 20030 & 55.67\% & 91.04\% & 76.09\% &  26980  &  18686  & 69.26\% & 97.03\%  & 99.74\% \\
    \textbf{SMART (QW2.5-7b)} & 35979 &  19971 & 55.51\% & 92.22\%   &  79.43\% & 26980 &  20375  & 75.52\% & 97.49\%  & 99.76\% \\
    \textbf{SMART (QW2.5-14b)} & 35979 &  22574  &  62.74\% & 93.71\% & 81.51\%  &  26980  &  21492 & 79.66\% &  98.37\%  & 99.79\% \\
    \textbf{SMART (QW2.5-32b)} & 35979 &  23083  & \cellcolor{gray!40}64.16\% &  93.80\%  & 83.32\% &  26980  & 20828  & 77.20\% &  98.41\%  & 99.72\% \\
    \hline
    \end{tabular}
    \\ {\scriptsize Note: The mutation generation approach achieving the best performance in terms of a metric is highlighted in a grey background color.}
  \tablabel{validity}
\end{table}

\tabref{validity} summarizes the validity results of different LLM-based mutation generation approaches. 
For each dataset, we report the expected number of mutations (\emph{Exp.}), the number of successfully parsed mutations (\emph{Gen.}), and three validity metrics: generation rate (\emph{Ge. Rate}), non-duplicate rate (\emph{ND. Rate}), and compilable rate (\emph{Com. Rate}).  
We note that GPT-4o is a proprietary model that does not support SFT. 
Therefore, for GPT-4o we only report the results of \ourtoolraw$_{RC}$, which integrates RAG and chunking but excludes fine-tuning.

Across both datasets, \ourtool consistently achieves higher validity than LLMut on all five open-source LLMs. 
For example, on Defects4J, SMART improves the generation rate of LLMut by more than 40 PPs on Qwen-2.5-14B (62.74\% vs. 22.40\%), while also maintaining a higher non-duplicate rate and compilability rate. 
Even under this restricted configuration, \ourtoolraw$_{RC}$ achieves substantial improvements over LLMut (GPT-4o) and LLMorpheus, achieving the highest non-duplicate rate and compilable rate on Defects4J.

On ConDefects, \ourtool yields generation rates above 60\% for all open-source models, compared to below 30\% for LLMut. 
Moreover, \ourtool achieves similar performance in terms of the non-duplicate rate and the compilability rate compared with LLMut with GPT-4o, and outperforms LLMorpheus with GPT-4o.

We also observe systematic differences between the two datasets. 
ConDefects generally exhibits much higher generation and compilability rates than Defects4J. 
This is largely due to the dataset characteristics: ConDefects tasks are algorithmic problems implemented as small single-file programs, which are structurally simpler and thus easier for LLMs to mutate while preserving syntactic correctness.


\begin{tcolorbox}[boxrule=1pt,left=2pt,right=2pt,top=2pt,bottom=2pt]
\underline{\textbf{Answer to RQ1:}} 
\ourtool substantially improves mutation validity. 
It increases the weighted average \textbf{generation rate} from 42.89\% (LLMut) to 65.6\% (+22.71 PPs; $\uparrow 52.95\%$). 
The \textbf{non-duplicate rate} rises to 95.62\% (+8.24–9.75 PPs over baselines), and the \textbf{compilable rate} reaches 90.21\% (up to +11.78 PPs). 
Moreover, all open-source LLMs achieve competitive performance with GPT-4o after applying \ourtool.
\end{tcolorbox}

\subsection{RQ2: Performance Comparison in the Effectiveness of Mutations}

\subsubsection{Metrics}
To evaluate the effectiveness of LLM-generated mutations, we adopt several complementary metrics that capture how well the generated mutations resemble real bugs. 
Following the definitions in Section~\ref{sec:back}, these metrics measure detectability, coupling, and semantic similarity:  
\begin{itemize}[leftmargin=*]
    \item \textbf{Real-Bug Detection (R.B.D.).}  
    The percentage of bug-revealing tests (i.e., tests that fail on the buggy version) that also kill at least one generated mutant.  
    This metric assesses whether the generated mutations can be detected by the same tests that expose real bugs.  

    \item \textbf{Coupling Rate (Coup.).}  
    The proportion of generated mutations that are coupled with real bugs, i.e., mutations that can be killed by at least one bug-revealing test.  
    This metric quantifies how closely mutations align with real-world defects.  

    \item \textbf{Average Ochiai Coefficient (Avg. O.).}  
    To further assess semantic similarity, we compute the Ochiai coefficient between each mutant and its corresponding real bug based on their sets of failing test cases.
    The calculation of \emph{Avg. O.} is shown in Formula~\ref{eq:aoc}.
    A higher score indicates stronger semantic similarity between generated mutations and real bugs.  
    
    \item \textbf{High-Similarity mutations (O. $\geq$ 0.8).}  
    Following Ojdanic et al.~\cite{ojdanic2023syntactic}, we also report the number of real bugs whose average Ochiai coefficient with the generated mutations is at least 0.8.  
    This stricter metric highlights how many real bugs are closely approximated by generated mutations.   
\end{itemize}
We also report the \textbf{Mutation Score (MS)} for completeness, but caution that it is not a sufficient measure of mutation generation effectiveness.  
A high mutation score may simply indicate that mutations are easy to kill, rather than being representative of real bugs.
Note that the closeness metric measures the semantic distance between a generated mutant and its corresponding historical real bug. Therefore, all baselines (including LLMut and LLMorpheus) use the same ground-truth bug as the reference.

\subsubsection{Results}

\begin{table}[tp]
  \centering
  \caption{The Effectiveness Comparison}
  \scriptsize
    \begin{tabular}{|l|l|l|l|l|l || l|l|l|l|l|}
    \hline
    \multirow{2}{*}{\textbf{Approaches}} & \multicolumn{5}{c||}{\textbf{Defects4J (701 Bugs)}} & \multicolumn{5}{c|}{\textbf{ConDefects (1290 Bugs)}} \\
\cline{2-11}       & \textbf{MS} & \textbf{R. B. D.} & \textbf{Coup.} & \textbf{Avg. O.} & \textbf{O.$\ge 0.8$} & \multicolumn{1}{l|}{\textbf{MS}} & \multicolumn{1}{l|}{\textbf{R. B. D.}} & \multicolumn{1}{l|}{\textbf{Coup.}} & \multicolumn{1}{l|}{\textbf{Avg. O.}} & \multicolumn{1}{l|}{\textbf{O.$\ge 0.8$}} \\
    \hline
    \textbf{LLMorpheus (GPT-4o)} & 74.20\% & 51.36\% & \cellcolor{gray!40}48.61\% & 16.14\% & 90 &  63.47\%  & 11.47\%   &  56.09\%  &  3.87\%  & 36 \\
    \hline
    \textbf{LLMut (GPT-4o)} & 67.78\%  &  29.91\%  &  37.89\%  &  10.37\%  &  90  &  79.71\%  &  7.36\%  &  63.29\%  &  3.79\%  & 66 \\
    \hline
    \textbf{LLMut (DS-6.7b)} & 70.71\%  & 25.25\%   &  45.82\%  &  8.55\%  &  60  &  79.34\%  & 30.00\%  &  65.07\%  & 15.67\%   & 203  \\
    \textbf{LLMut (LM3.1-8b)} &  58.23\%  &  22.54\%  & 32.09\%  &  5.53\%  &  33  &  79.86\%  &  19.15\%  &  66.85\%  &  9.51\%  &  123 \\
    \textbf{LLMut (QW2.5-7b)} &  59.95\%  & 44.51\%   & 36.53\%   & 14.58\%   &  103  &  81.80\%  &  22.64\%  &  66.32\%  & 11.24\%   & 188 \\
    \textbf{LLMut (QW2.5-14b)} &  65.15\%  &  58.92\%  &  39.78\%  &  22.69\%  &  205  &  80.67\%  &  65.58\%  & 66.31\%  & 35.15\%    &  576 \\
    \textbf{LLMut (QW2.5-32b)} &  66.51\%  & 77.75\%  & 37.32\%   &   25.26\% &  278  & 81.24\%   &  80.62\%  &  66.04\%  & 43.01\%   & 730 \\
    \hline
    \textbf{SMART$_{RC}$ (GPT-4o)} &  69.41\%  &  89.44\%  &  37.24\%  &  27.61\% &  288  &   76.93\% &  78.76\%  &  62.94\%  &   41.18\% & 530 \\
    \hline
    \textbf{SMART (DS-6.7b)} &  74.72\%  &  92.72\%  &  40.21\%  & 28.70\%   &  298  &  84.42\%  &  95.74\%  &  69.51\%  & \cellcolor{gray!40}51.05\%   & \cellcolor{gray!40}739  \\
    \textbf{SMART (LM3.1-8b)} & 65.46\%   &  91.16\%  &  37.54\%  &  27.15\%  &  250  &  84.62\%  &  93.41\%  &  70.66\%  & 49.41\%   & 657  \\
    \textbf{SMART (QW2.5-7b)} &  \cellcolor{gray!40}76.12\%  &  \cellcolor{gray!40}93.58\%  &  43.04\%  &  \cellcolor{gray!40}29.30\%  &  275  &  85.56\%  &  94.19\%  &  72.16\%  &  50.12\%  & 697 \\
    \textbf{SMART (QW2.5-14b)} &  75.76\%  &  93.87\%  & 42.14\%   &  29.23\%  &  289  &  87.32\%  &  94.26\%  &  72.23\%  & 50.22\%   & 709 \\
    \textbf{SMART (QW2.5-32b)} & 75.99\%  &  92.87\%  &  42.77\%  &  29.16\%  &  \cellcolor{gray!40}300  & \cellcolor{gray!40}87.44\% &  \cellcolor{gray!40}94.50\%  &  \cellcolor{gray!40}72.37\%  & 49.63\%   & 712 \\
    \hline
    \end{tabular}
    \\ {\scriptsize Note: The mutation generation approach achieving the best performance in terms of a metric is highlighted in a grey background color.}
  \tablabel{effectiveness}
\end{table}

\tabref{effectiveness} presents the results of the effectiveness evaluation on both Defects4J and ConDefects. 
We can make the following key observations.

First, across nearly all metrics, \ourtool significantly outperforms the baseline approaches.  
On Defects4J,  \ourtool (Qwen-2.5-7b) achieves the best mutation score (76.12\%) and the highest real-bug detection rate (93.58\%), exceeding the strongest LLMut variant by more than 15 PPs.  
Similarly, \ourtool (Qwen-2.5-32b) yields the largest number of high-Ochiai bugs (300), indicating its ability to generate mutations that are strongly coupled with real bugs.  
In terms of semantic similarity, \ourtool with different models obtains higher average Ochiai coefficients (around 27--29\%) than their LLMut counterparts (mostly below 23\%).

Second, the advantage of \ourtool is even more pronounced on ConDefects.  
For example, \ourtool (DS-6.7b) reaches an average Ochiai coefficient of 51.05\%, compared to only 15.67\% for LLMut (DS-6.7b).  
Likewise, \ourtool (Qwen-2.5-14b) and \ourtool (Qwen-2.5-32b) achieve the highest real-bug detection rates (above 94\%) and coupling rates (above 72\%), along with over 700 high-Ochiai bugs, far surpassing all baseline models.  

Third, while LLMorpheus (GPT-4o) shows relatively high mutation scores (74.20\% on Defects4J), it performs poorly in semantic similarity (16.14\% Avg. O.) and high-Ochiai bug count (90).  
In contrast, \ourtoolraw$_{RC}$ (GPT-4o), even without fine-tuning, improves the average Ochiai to 27.61\% and detects 288 high-Ochiai bugs. 

Finally, by comparing \ourtoolraw$_{RC}$ (GPT-4o) and the full \ourtool with open-source LLMs, we observe that \textbf{fine-tuning enables 7B-scale models to achieve performance comparable to, and in some cases even surpassing, GPT-4o.}
This finding highlights the practical value of our approach, as it demonstrates that competitive effectiveness can be attained even with relatively small open-source models.

Overall, these results demonstrate that \ourtool consistently generates more effective mutations than both LLMut and LLMorpheus.  
By improving real-bug detection, fault coupling, and semantic similarity, \ourtool provides a stronger foundation for mutation-based testing applications.

\begin{tcolorbox}[boxrule=1pt,left=2pt,right=2pt,top=2pt,bottom=2pt]
\underline{\textbf{Answer to RQ2:}} \ourtool consistently improves the effectiveness of LLM-generated mutations compared to LLMut and LLMorpheus. 
For the \textbf{real bug detection rate}, \ourtool achieves a weighted average of 92.61\%, compared to 57.86\% for LLMut ($\uparrow 60.1\%$; +34.75 PPs) and 31.99\% for LLMorpheus ($\uparrow 189.5\%$; +60.62 PPs).  
For the \textbf{coupling rate}, \ourtool reaches 54.94\%, which is slightly higher than LLMut (53.12\%) and LLMorpheus (52.23\%).  
For the \textbf{average Ochiai coefficient}, \ourtool achieves 38.44\%, compared to 25.61\% for LLMut ($\uparrow 50.1\%$; +12.83 PPs) and 10.18\% for LLMorpheus ($\uparrow 277.6\%$; +28.26 PPs).  
These results confirm that \ourtool generates mutants with stronger alignment to real-world faults, leading to substantial improvements in mutation effectiveness.
Moreover, through \ourtool, fine-tuning elevates 7B-scale models to a level on par with GPT-4o.
\end{tcolorbox}

\subsection{RQ3: Performance Comparison on Mutation-Based Test Case Prioritization}

\subsubsection{Metrics}
In mutation-based test case prioritization (TCP), the quality of generated mutations is indirectly reflected by the prioritization effectiveness achieved when they are used as mutation faults.  
Following prior work~\cite{do2006use,tian2022learning,tianleam++}, we adopt three widely studied mutation-based TCP techniques, \textbf{GRK} (Greedy Kill), \textbf{GRD} (Greedy Distinguish), and \textbf{HYB-$\omega$}, which have been introduced in Section~\ref{sec:mbtcp}.
For each TCP technique, we measure effectiveness using the \textbf{Average Percentage of Faults Detected (APFD)}~\cite{rothermel1999test}, a standard metric in TCP research, computed as follows:
\begin{equation}
    APFD = 1- \frac{ TF_1 + TF_2 + \dots + TF_n }{ n r } + \frac{1}{2n} ,
\end{equation}
where $r$ refers to the number of bugs, $n$ refers to the number of test cases to be prioritized, and $TF_i$ refers to the ranking of the first test case prioritized by a TCP approach that detects the $i^{th}$ bug.
A higher APFD value indicates that real bugs are detected earlier during the prioritized execution of test cases.

\subsubsection{Results}
\tabref{mbtcp} reports the APFD results of three mutation-based TCP techniques (GRK, GRD, HYB-$\omega$) when using mutations generated by different approaches.

On Defects4J, \ourtool consistently outperforms LLMut and LLMorpheus across all three TCP techniques.
For example, with Qwen-2.5-32B, \ourtool attains the highest APFDs: 0.867 (GRK), 0.8805 (GRD), and 0.8805 (HYB-$\omega$), exceeding the corresponding LLMut scores (0.8181/0.8312/0.8312).
Compared with LLMut with the same model as \ourtool, \ourtool consistently outperforms on all metrics, showing the benefit of chunking, RAG, and fine-tuning.
On ConDefects, \ourtool is competitive but does not surpass the best LLMut variants.  
The top scores are achieved by LLMut (Qwen-2.5-32B) for GRK (0.8823) and by LLMut (Qwen-2.5-14B) for GRD/HYB-$\omega$ (0.8684).  
\ourtool reaches 0.869 (GRK, DeepSeek-6.7B) and 0.865 (GRD/HYB-$\omega$, DeepSeek-6.7B), within 1.33 percentage points of the best LLMut results, and remains above LLMorpheus and LLMut (GPT-4o).

\begin{table}[tp]
  \centering
  \scriptsize
  \caption{Mutation-Based Test Case Prioritization}
    \begin{tabular}{|l|l|l|l || l|l|l|}
    \hline
    \multicolumn{1}{|c|}{\multirow{2}{*}{\textbf{Approaches}}} & \multicolumn{3}{c||}{\textbf{Defects4J (701 Bugs)}} & \multicolumn{3}{c|}{\textbf{ConDefects (1290 Bugs)}} \\
\cline{2-7}       & \textbf{GRK} & \textbf{GRD} & \textbf{HYB-$\omega$} & \textbf{GRK} & \textbf{GRD} & \textbf{HYB-$\omega$} \\
    \hline
    \textbf{LLMorpheus (GPT-4o)} & 0.6707   &  0.6746  & 0.6746   &  0.8445  & 0.844   & 0.844 \\
    \hline
    \textbf{LLMut (GPT-4o)} &  0.6246  & 0.6295   &  0.6295  &  0.8521  &  0.8649  & 0.8649 \\
    \hline
    \textbf{LLMut (DS-6.7b)} &  0.6  &  0.5995  &  0.5995  &  0.8509  &  0.8523  & 0.8523 \\
    \textbf{LLMut (LM3.1-8b)} &  0.565  &  0.5657  &  0.5657  &  0.8471  &  0.8398  & 0.8398 \\
    \textbf{LLMut (QW2.5-7b)} &  0.6736  &  0.677  &  0.677  &  0.8551  &  0.8637  & 0.8637 \\
    \textbf{LLMut (QW2.5-14b)} &  0.7436  &  0.7493  &  0.7493  & 0.8768   &  \cellcolor{gray!40}0.8684  & \cellcolor{gray!40}0.8684 \\
    \textbf{LLMut (QW2.5-32b)} &  0.8181  &  0.8312  & 0.8312   &  \cellcolor{gray!40}0.8823  & 0.876   & 0.876 \\
    \hline
    \textbf{SMART$_{RC}$ (GPT-4o)} &  0.8399  &  0.8552  &  0.8552  &  0.8675  &  0.8623  & 0.8623 \\
    \hline
    \textbf{SMART (DS-6.7b)} &  0.8522  & 0.8793   &  0.8793  &  0.869  &  0.865  & 0.865 \\
    \textbf{SMART (LM3.1-8b)} &  0.8285  &  0.849  &  0.849  &  0.8585  &  0.8523  &  0.8523 \\
    \textbf{SMART (QW2.5-7b)} &  0.8442  &  0.8635  &  0.8635  &  0.8658  &  0.8636  & 0.8636 \\
    \textbf{SMART (QW2.5-14b)} &  0.8563  & 0.8735   &  0.8735  &  0.8595  & 0.8577   &  0.8577 \\
    \textbf{SMART (QW2.5-32b)} &  \cellcolor{gray!40}0.867  &  \cellcolor{gray!40}0.8805  &  \cellcolor{gray!40}0.8805  &  0.8658  &  0.8609  & 0.8609 \\
    \hline
    \end{tabular}
  \tablabel{mbtcp}
\end{table}

\begin{tcolorbox}[boxrule=1pt,left=2pt,right=2pt,top=2pt,bottom=2pt]
\underline{\textbf{Answer to RQ3:}} On Defects4J, \ourtool yields state-of-the-art APFD across GRK/GRD/HYB-$\omega$, clearly outperforming LLMut and LLMorpheus; on ConDefects, it remains highly competitive (within 0.3--1.3 PPs of the best LLMut variants) while \ourtoolraw$_{RC}$ (GPT-4o) also surpasses prior GPT-4o baselines.
\end{tcolorbox}

\subsection{RQ4: Performance Comparison on Mutation-based Fault Localization}

\subsubsection{Metrics}
In mutation-based fault localization (MBFL), we evaluate the quality of generated mutations by examining their impact on the accuracy of localizing buggy statements.  
Given a buggy method, each statement is ranked based on its suspiciousness score computed from mutation analysis (Section~\ref{sec:mbfl}).  
We then measure the effectiveness of fault localization using the following standard metrics:  

\begin{itemize}[leftmargin=*]
    \item \textbf{Top-$k$.}  
    The proportion of bugs for which at least one faulty line is ranked within the top $k$ positions.  
    Higher Top-$k$ values indicate that faulty lines are placed earlier in the ranking, reducing the manual effort needed for debugging. In our study, we set $k$ to 1, 3, and 5, following the common practice of MBFL research~\cite{tian2022learning}.

    \item \textbf{Mean Average Rank (MAR).}  
    For each fault, we record the rank position of the first faulty line and average this value across all bugs.  
    A lower MAR indicates that bugs are ranked closer to the top on average, thus easier to locate.  

    \item \textbf{Mean First Rank (MFR).}  
    Similar to MAR, but computed as the mean of the first relevant rank across all bugs.  
    This metric emphasizes the effort required to examine the ranked list until the first faulty statement is reached.  
\end{itemize}
If multiple statements share the same suspiciousness score, we follow prior work~\cite{zou2019empirical,tian2022learning} and use $E_{inspect}$ to compute their expected rank.
In MBFL, when the mutations produced by one technique lead to larger Top-$k$ scores and smaller MFR or MAR values compared to another, the former is regarded as achieving better fault localization performance.

\subsubsection{Results}
As shown in \tabref{mfbl-d4j} and \tabref{mfbl-cond}, \ourtool consistently improves the usefulness of mutations for MBFL on both MUSE and Metallaxis. 
First, \ourtool yields higher Top-$k$ accuracies than LLMut and LLMorpheus on both Defects4J and ConDefects.
For instance, in terms of Top-1, \ourtool (Qwen-2.5-7B) localizes 134 bugs by MUSE and 119 bugs by Metallaxis on Defects4J, and localizes 132 bugs by MUSE and 115 bugs on ConDefects, significantly outperforming LLMut and LLMorpheus.
Second, \ourtool also reduces ranking effort: on both Defects4J and ConDefects, \ourtool (DS-6.7B) achieves the lowest MAR/MFR values.
These results suggest that mutations generated by \ourtool not only push buggy lines into the top ranks more frequently but also improve overall localization quality across datasets.

\begin{table}[tp]
  \centering
  \scriptsize
  \caption{Statement-Level Mutation-Based Fault Localization on Defects4J (701 Bugs)}
    \begin{tabular}{|l|l|l|l|l|l|l|l|l|l|l|}
    \hline
    \multicolumn{1}{|c|}{\multirow{2}{*}{\textbf{Approaches}}} & \multicolumn{5}{c|}{\textbf{MUSE}} & \multicolumn{5}{c|}{\textbf{Metallaxis}} \\
\cline{2-11}       & \textbf{Top-1} & \textbf{Top-3} & \textbf{Top-5} & \textbf{MAR} & \textbf{MFR} & \textbf{Top-1} & \textbf{Top-3} & \textbf{Top-5} & \textbf{MAR} & \textbf{MFR} \\
    \hline
    \textbf{LLMorpheus (GPT-4o)} &   100 &  143  &  148  &  157.61  & 157.52   & 93   & 144   &  148  &  157.58  & 157.52 \\
    \hline
    \textbf{LLMut (GPT-4o)} &  56  &  119  &  149  &  143.73  &  143.11  &  49  &  126  & 150  &  143.88  & 143.16 \\
    \hline
    \textbf{LLMut (DS-6.7b)} &  57 &  118  & 151  &  151.56 & 151.30   &  55  & 118 & 148  &  151.60 & 151.33 \\
    \textbf{LLMut (LM3.1-8b)} &  69  &  111  & 127   &  162.25  &  162.16  &  68  & 111 &  127  &  162.21  & 162.15 \\
    \textbf{LLMut (QW2.5-7b)} & 85  & 198  &  247  &  119.88  &  119.32  & 78   &  201  & 240   &  119.96  & 119.34 \\
    \textbf{LLMut (QW2.5-14b)} & 104   &  252  & 308  &  94.56 & 93.62   & 97  & 259  &  309  &  94.58  & 93.59 \\
    \textbf{LLMut (QW2.5-32b)} & 107  &  280  &  365  & 59.55   & 57.36   &  93	&  292 &  376  &  57.12  & 55.10 \\
    \hline
    \textbf{SMART$_{RC}$ (GPT-4o)} & 105 &  293 & 360  &  59.72  &  57.39  &  92  &  299  &  374  &  60.13  & 57.34 \\
    \hline
    \textbf{SMART (DS-6.7b)} &  116  &  306	  & \cellcolor{gray!40}397  &  \cellcolor{gray!40}52.64  & \cellcolor{gray!40}50.91 &  97  &  \cellcolor{gray!40}316  &  \cellcolor{gray!40}399  &  \cellcolor{gray!40}52.93  & \cellcolor{gray!40}50.94 \\
    \textbf{SMART (LM3.1-8b)} &  78  &  251	 &  327  &  76.40 &  75.29  & 76   &  260  &  332  &  76.51  &  75.16 \\
    \textbf{SMART (QW2.5-7b)} &  \cellcolor{gray!40}134  & \cellcolor{gray!40}314	 &  392 &  59.68  &  58.27  &  \cellcolor{gray!40}119  &  314  &  390  &  59.87  & 58.27 \\
    \textbf{SMART (QW2.5-14b)} &  130 & 303  &  378  &  56.79 &  55.09  & 111   & 310  & 383   & 57.12   & 55.10 \\
    \textbf{SMART (QW2.5-32b)} &  129  & 309   & 377   &  56.52  &  54.78  &  116  &  305  & 386  &  56.77  & 54.77 \\
    \hline
    \end{tabular}
  \tablabel{mfbl-d4j}
\end{table}

\begin{table}[tp]
  \centering
  \scriptsize
  \caption{Statement-Level Mutation-Based Fault Localization on ConDefects (1290 Bugs)}
    \begin{tabular}{|l|l|l|l|l|l|l|l|l|l|l|}
    \hline
    \multicolumn{1}{|c|}{\multirow{2}{*}{\textbf{Approaches}}} & \multicolumn{5}{c|}{\textbf{MUSE}} & \multicolumn{5}{c|}{\textbf{Metallaxis}} \\
\cline{2-11}       & \textbf{Top-1} & \textbf{Top-3} & \textbf{Top-5} & \textbf{MAR} & \textbf{MFR} & \textbf{Top-1} & \textbf{Top-3} & \textbf{Top-5} & \textbf{MAR} & \textbf{MFR} \\
    \hline
    \textbf{LLMorpheus (GPT-4o)} & 11   &  16  &  16  &  197.53  &  197.53  & 12 &  16  &  16 &  197.53  & 197.53 \\
    \hline
    \textbf{LLMut (GPT-4o)} &  10  &  35  & 53  & 186.75   &  186.74  & 8  &  34  &  51 &  186.79  & 186.78 \\
    \hline
    \textbf{LLMut (DS-6.7b)} & 24  &  105 & 198 &  148.39  & 148.39  & 28  &  117  &  193 &  148.42  & 148.42 \\
    \textbf{LLMut (LM3.1-8b)} &  14	&  45 & 72  &  176.29  &  176.28  & 15  &  47  &  69 &  176.35 & 176.34 \\
    \textbf{LLMut (QW2.5-7b)} &  20	 &  56  &  95  &  163.12  &  163.12  & 17  &  60 &  95 & 163.27 & 163.27 \\
    \textbf{LLMut (QW2.5-14b)} &  103  &  263  &  383  &  88.09  &  88.08  & 79	 &  246  &  373 &  88.41  & 88.39 \\
    \textbf{LLMut (QW2.5-32b)} &  106  & 308  &  475 & 58.26   &  58.25  & 85  & 283  & 450  & 58.79   & 58.77 \\
    \hline
    \textbf{SMART$_{RC}$ (GPT-4o)} &  60  &  225  & 350  &  114.35  & 114.35   & 61	  &  221  &  335 &  114.55  & 114.54 \\
    \hline
    \textbf{SMART (DS-6.7b)} &  105  & 349  & 499   &  \cellcolor{gray!40}70.16  &  \cellcolor{gray!40}69.96  & 102  &  340  & 479  & \cellcolor{gray!40}70.54 & \cellcolor{gray!40}70.35 \\
    \textbf{SMART (LM3.1-8b)} & 120	 & 287   &  398 & 115.08  & 115.02  &  101 & 297  & 392  & 115.17  & 115.09 \\
    \textbf{SMART (QW2.5-7b)} & 132	 &  337  &  474  &  101.36  &  101.28 & 115 & 336 &  454 &  101.48  & 101.39 \\
    \textbf{SMART (QW2.5-14b)} & \cellcolor{gray!40}143   &  349  &  476  &  90.03  &  89.89  &  \cellcolor{gray!40}120 & 340   & 464  &  90.31  & 90.16 \\
    \textbf{SMART (QW2.5-32b)} &  132  &  \cellcolor{gray!40}367  & \cellcolor{gray!40}516  & 85.33  & 85.18  &  109 & \cellcolor{gray!40}354   &  \cellcolor{gray!40}499 &  85.58  & 85.39 \\
    \hline
    \end{tabular}
  \tablabel{mfbl-cond}
\end{table}

\begin{tcolorbox}[boxrule=1pt,left=2pt,right=2pt,top=2pt,bottom=2pt]
\underline{\textbf{Answer to RQ4:}} \ourtool enhances MBFL by (1) achieving higher Top-$k$ accuracies than LLMut and LLMorpheus on both datasets, and (2) reducing MAR/MFR on Defects4J and remaining competitive on ConDefects.
For instance, on Defects4J, \ourtool (QW-14b) ranks 143 bugs as Top-1 by MUSE (LLMut has 106, LLMorpheus has 11), while \ourtool (DS-6.7b) achieves 399 bugs ranked with Top-5 by Metallaxis (LLMut has 376, LLMorpheus has 148).
Thus, \ourtool generates mutations that more effectively support fault localization by ranking buggy lines earlier.
\end{tcolorbox}

\subsection{RQ5: Ablation Study on Components}

\figref{radar} presents radar charts that summarize representative metrics across the first four RQs, including validity, effectiveness, MBTCP (GRK), and MBFL (MUSE Top-5 and Metallaxis Top-5).
We observe that enabling all three optimizations in \ourtool (RAG, chunking, and fine-tuning) achieves the best overall performance across different LLMs.
Among the components, chunking provides the most significant improvements.
RAG, on the other hand, shows substantial gains when applied alone (i.e., \ourtoolraw$_{R}$ vs. LLMut).
However, once chunking is enabled, the incremental benefits of RAG diminish, suggesting that chunking already provides sufficient contextual guidance.
Fine-tuning further adapts mid-scale open-source models (7B–14B) to mutation generation, allowing them to reach or even surpass GPT-4o under the full \ourtool setup.

\begin{figure}[t]
  \centering
  
  \begin{subfigure}[t]{0.30\linewidth}
    \centering
    \includegraphics[width=\linewidth]{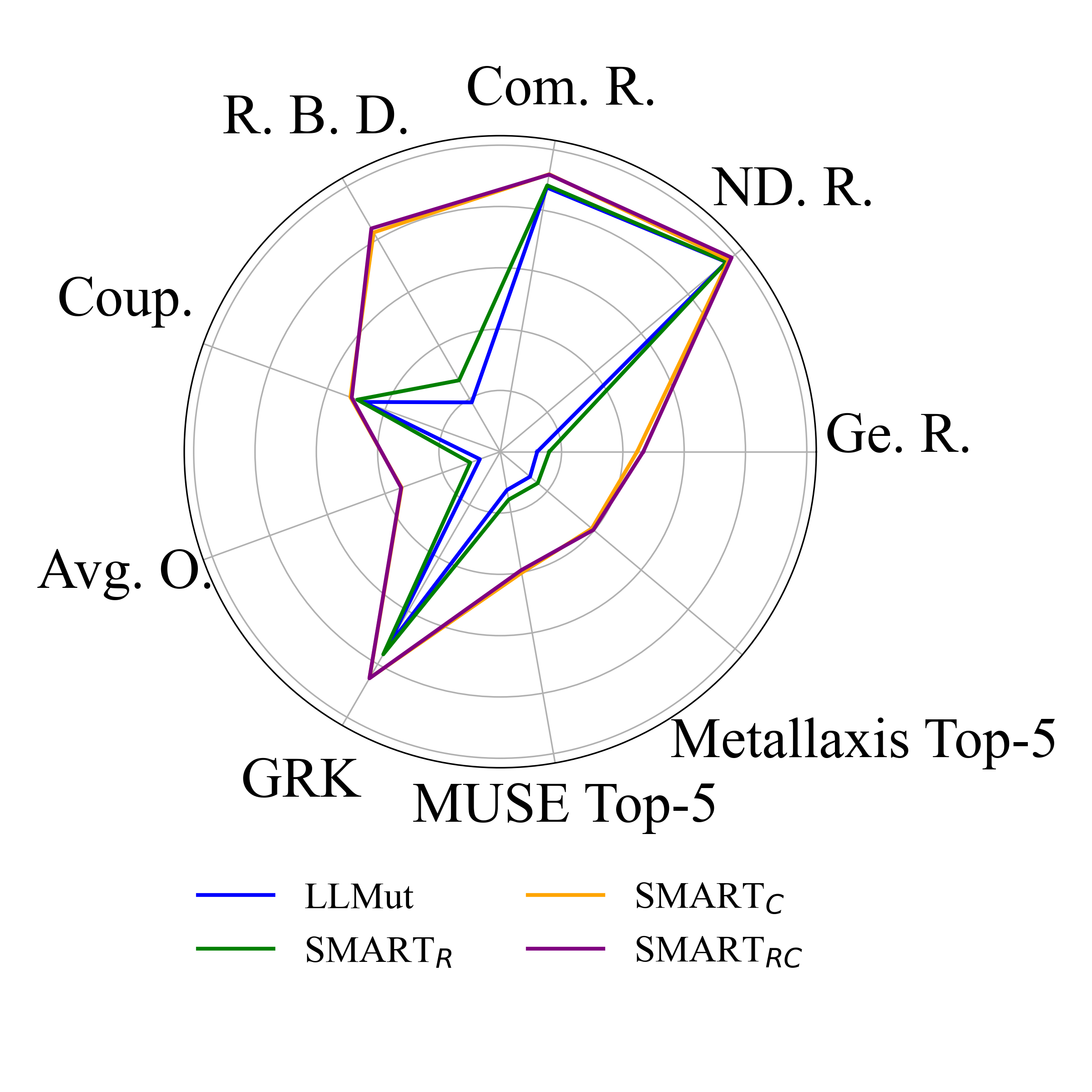}
    \caption{\ourtool (GPT-4o)}
    \figlabel{radar-gpt}
  \end{subfigure}
  \begin{subfigure}[t]{0.30\linewidth}
    \centering
    \includegraphics[width=\linewidth]{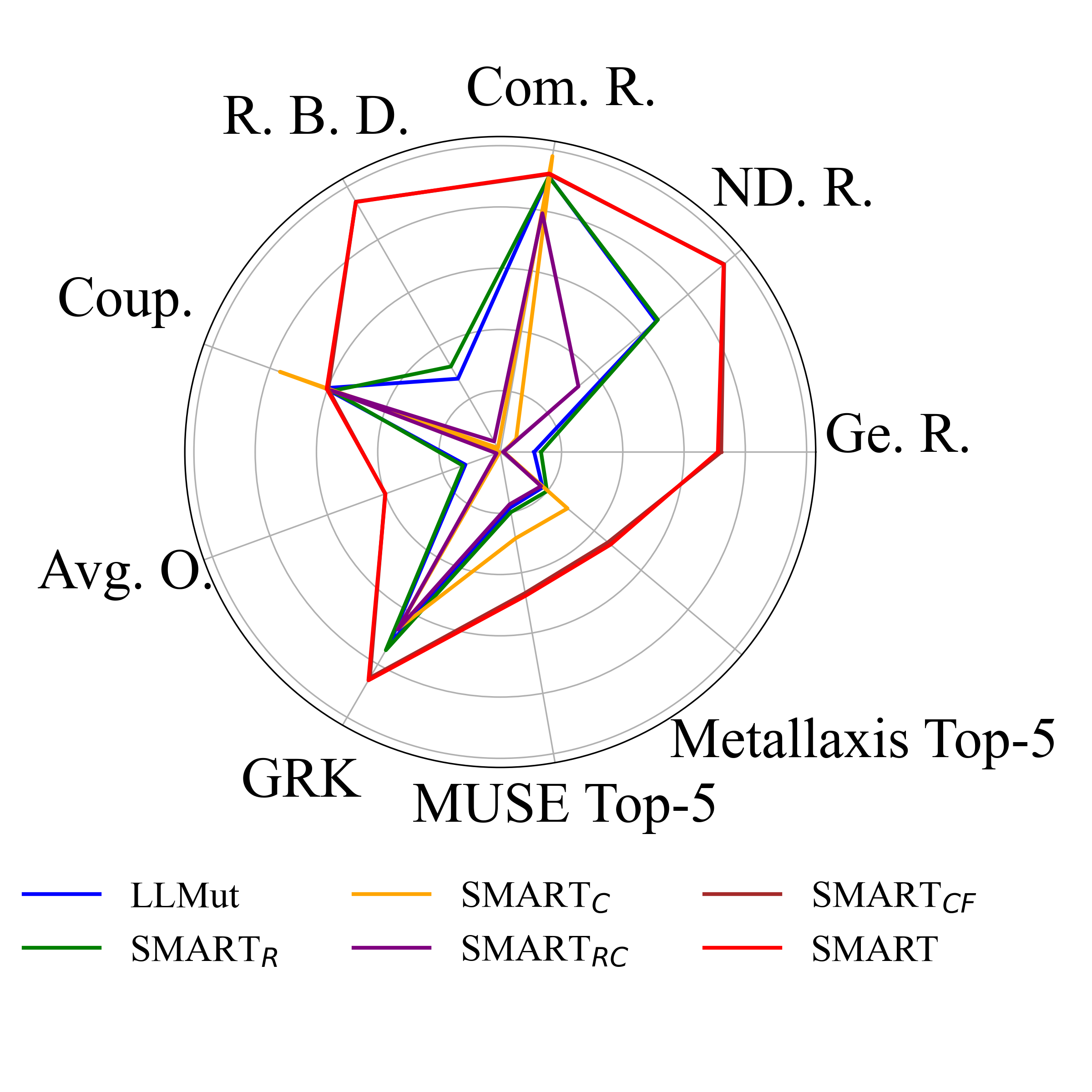}
    \caption{\ourtool (DS-6.7b)}
    \figlabel{radar-ds}
  \end{subfigure}
  \begin{subfigure}[t]{0.30\linewidth}
    \centering
    \includegraphics[width=\linewidth]{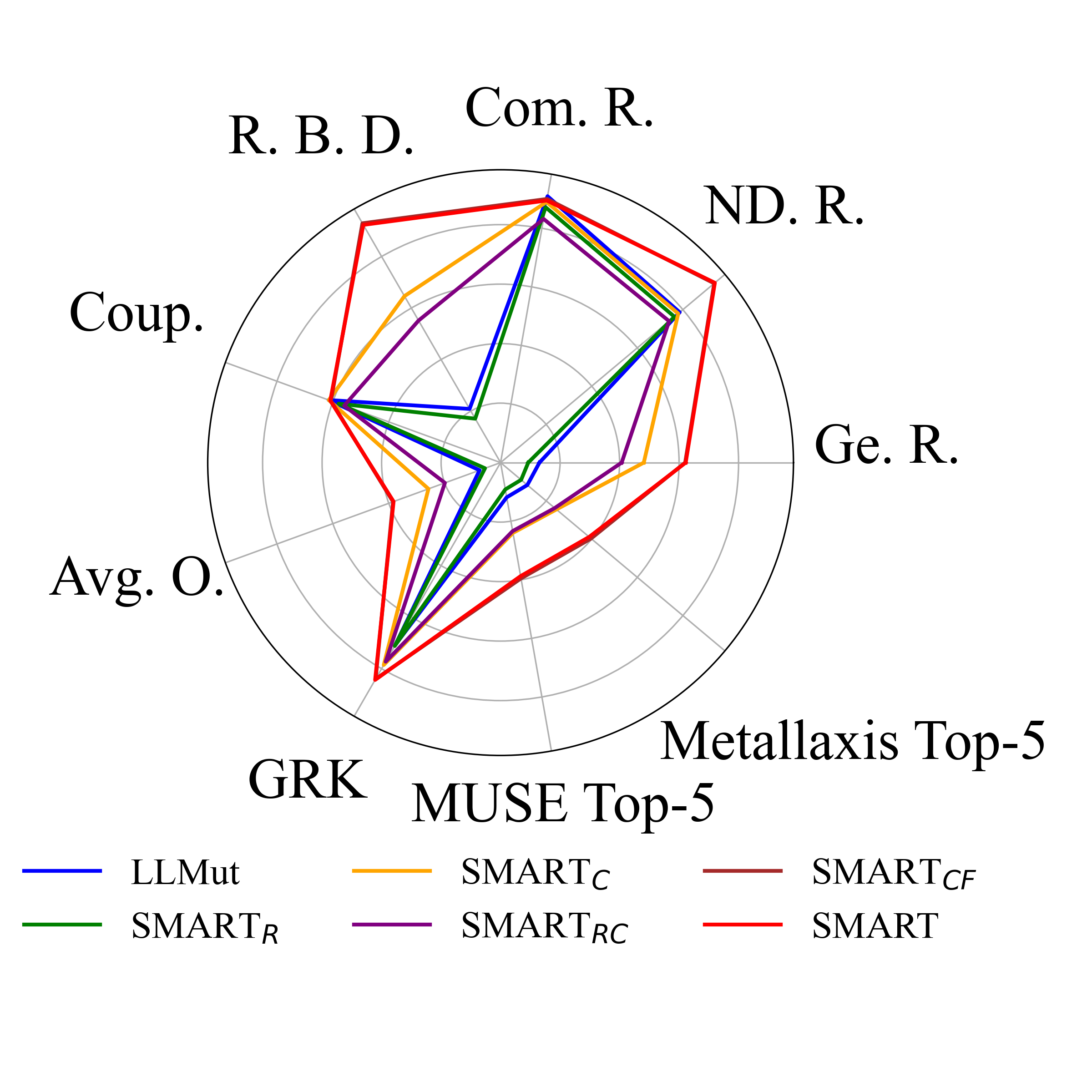}
    \caption{\ourtool (LM3.1-8b)}
    \figlabel{radar-lm}
  \end{subfigure}

  \begin{subfigure}[t]{0.30\linewidth}
    \centering
    \includegraphics[width=\linewidth]{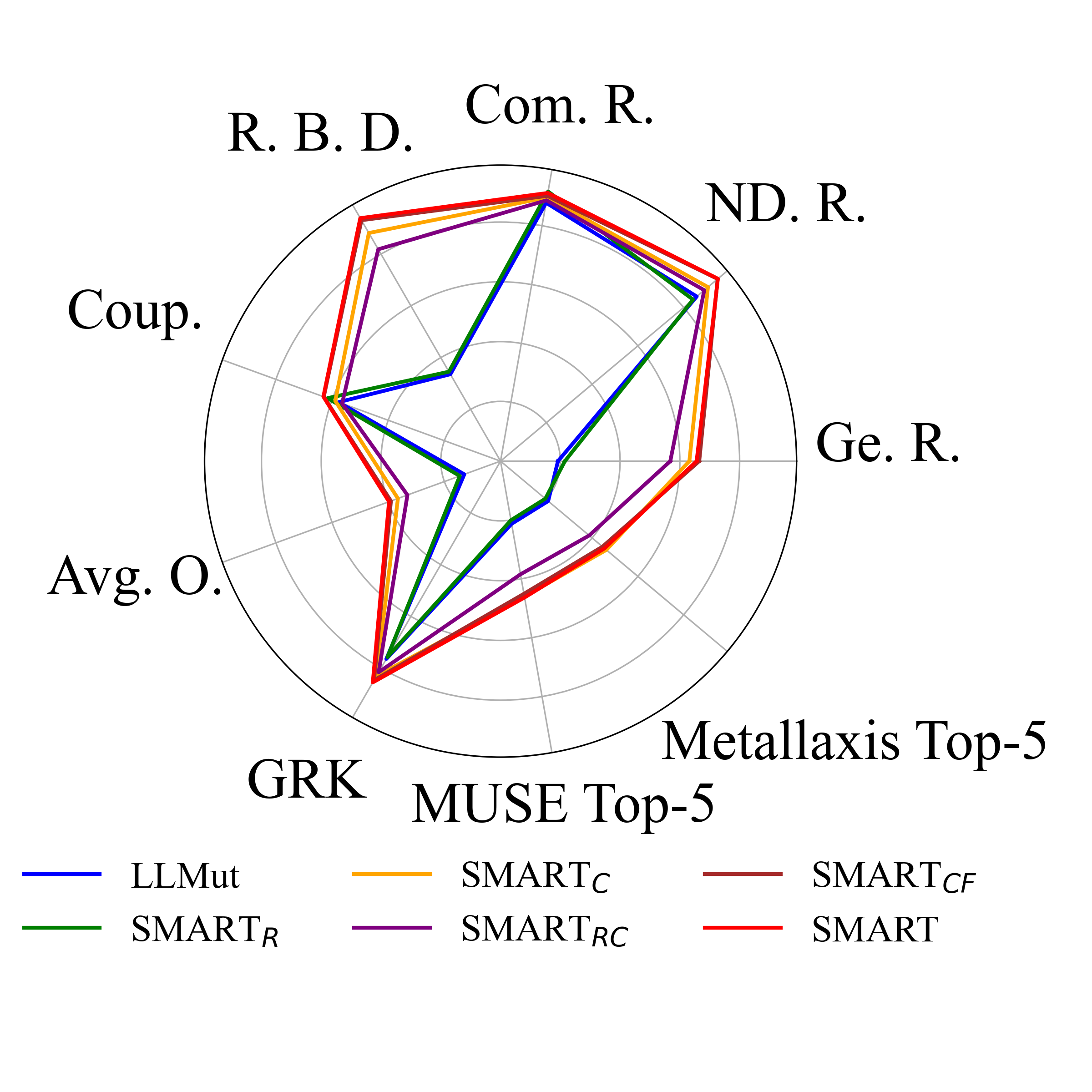}
    \caption{\ourtool (QW2.5-7b)}
    \figlabel{radar-qw-7}
  \end{subfigure}
  \begin{subfigure}[t]{0.30\linewidth}
    \centering
    \includegraphics[width=\linewidth]{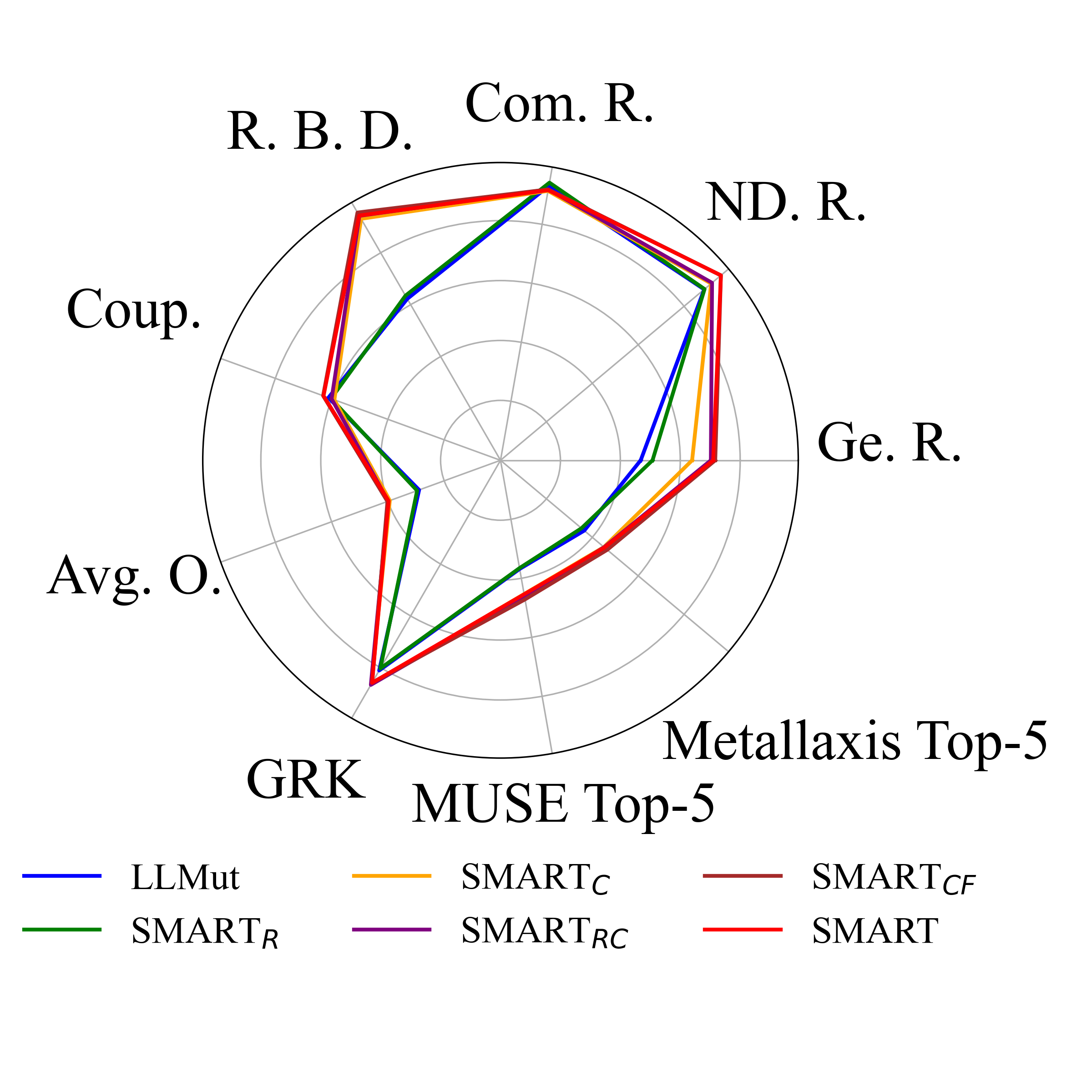}
    \caption{\ourtool (QW2.5-14b)}
    \figlabel{radar-qw-14}
  \end{subfigure}
  \begin{subfigure}[t]{0.30\linewidth}
    \centering
    \includegraphics[width=\linewidth]{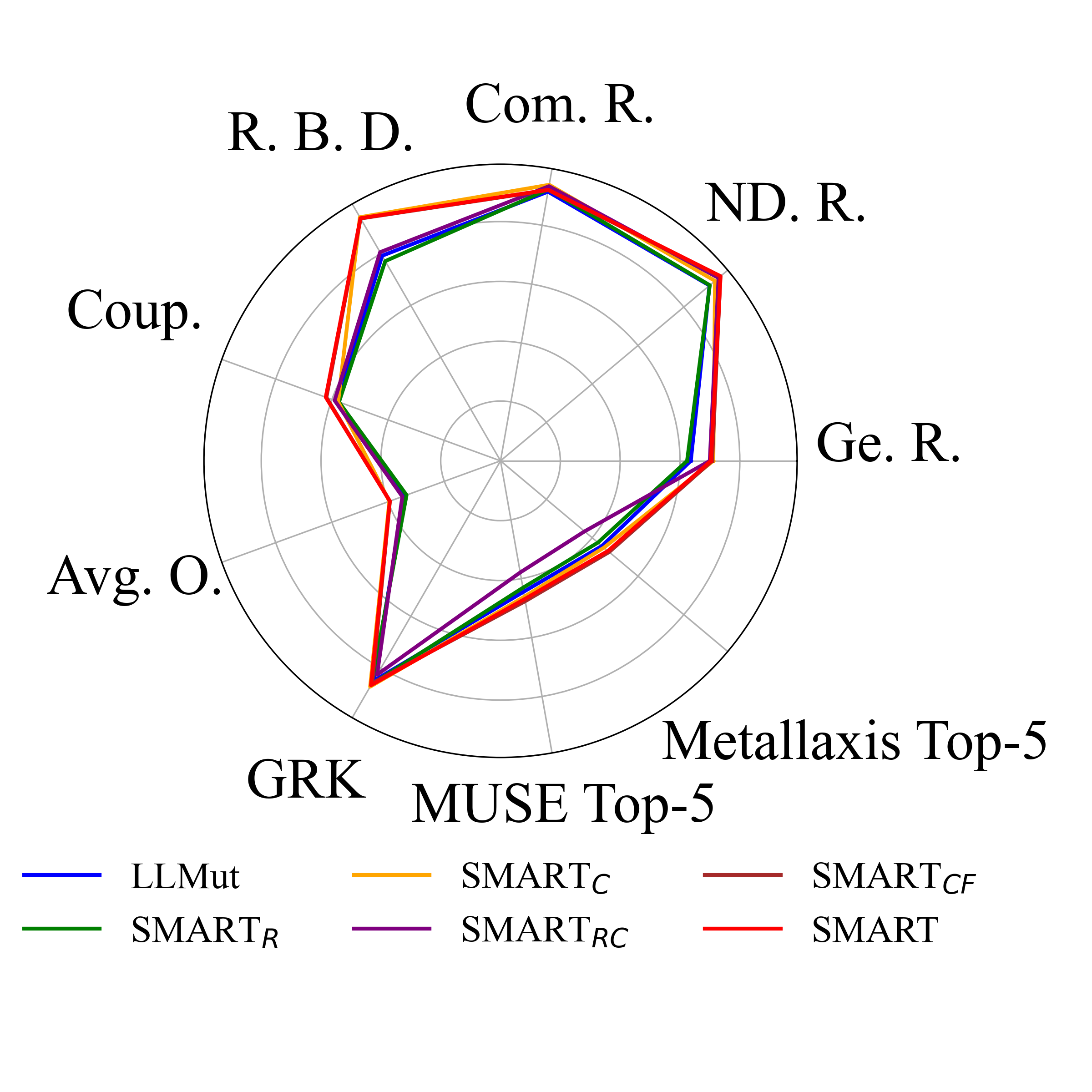}
    \caption{\ourtool (QW2.5-32b)}
    \figlabel{radar-qw-32}
  \end{subfigure}
  
   \caption{Comparison among Different Variants of \ourtool}
  \figlabel{radar}
\end{figure}

\begin{tcolorbox}[boxrule=1pt,left=2pt,right=2pt,top=2pt,bottom=2pt]
\underline{\textbf{Answer to RQ5:}} All three components of \ourtool---RAG, chunking, and fine-tuning---contribute to improving mutation generation and its downstream effectiveness.
Among them, code chunking consistently brings the most significant gains.
\end{tcolorbox}

\section{Discussion} \label{sec:diss}

\subsection{Token Cost for Mutation Generation}


\begin{table}[tp]
  \centering
  \scriptsize
  \caption{The Statistics of Token Costs}
    \begin{tabular}{l|l|l|l|l}
    \hline
       & \textbf{LLMut} & \textbf{\ourtoolraw$_{R}$} & \textbf{\ourtoolraw$_{C}$} & \textbf{\ourtoolraw} \\
    \hline
    \textbf{Avg. Query Time per Mtd.} & 1  & 1  & 4.6 & 4.6 \\
    \textbf{Avg Tokens per Query} & 716 & 778 & 950 & 979 \\
    \hline
    \end{tabular}
  \tablabel{cost}
\end{table}

While \ourtool improves the quality and effectiveness of generated mutants, it also introduces additional token and time costs due to its design choices. 
Both RAG and code chunking increase prompt size compared with LLMut. 
RAG adds retrieved bug–fix examples into the prompt, while chunking transforms a single query at the method level into multiple queries at the chunk level. 
As a result, chunking amplifies the total token consumption and query time.

\tabref{cost} reports the average query time and token usage per method. 
Compared with LLMut, RAG alone (\ourtoolraw$_R$) only increases average tokens per query from 716 to 778 due to the inclusion of few-shot examples.
By contrast, chunking (\ourtoolraw$_C$) substantially increases the number of LLM invocations, leading to an average of 4.6 queries per method and a token cost of 950 tokens per query. 
The full \ourtool has the same average of queries as chunking alone, while it costs 979 tokens per query.
In summary, \ourtool incurs higher token usage primarily due to chunking. 
Overall, the average token cost per query increases by about \textbf{36.7\%} (from 716 in LLMut to 979 in \ourtool).


\subsection{Cost Analysis of \ourtool}

For the GPU-related steps of \ourtool, which dominate the overall cost, we report the \textbf{wall-clock time} to capture its end-to-end execution cost. Moreover, for the SFT stage, which is the potential hardware bottleneck for deploying LLM-based approaches, we further report two additional metrics: \textbf{peak GPU memory usage} and \textbf{floating-point operations (FLOPs)}. Peak GPU memory usage reflects the practical hardware requirement and determines whether the fine-tuning process can be executed on commodity GPUs. FLOPs quantify the total amount of numerical computation performed during training, providing a hardware-independent measure of computational intensity. To mitigate the effect of randomness, we repeat each experiment three times and report the average results.

\begin{table}[tp]
  \centering
  \scriptsize
  \caption{RAG Related-costs of \ourtool}
    \begin{tabular}{l|r||r|r||r|r}
    \hline
       & \textbf{RAG Building} & \textbf{RAG Retrieval (Mtd)} & \textbf{Avg. Mtd} & \textbf{RAG Retrieval (Chk)} & \textbf{Avg. Chk} \\
    \hline
    \textbf{Time (s)} & 613 & 5,468 & 2.7 & 12,961 & 6.5 \\
    \hline
    \end{tabular}
  \tablabel{rag-cost}
\end{table}

\tabref{rag-cost} reports the wall-clock time for RAG building and retrieval.
RAG building takes 613 seconds, which is a one-time offline cost for constructing the embedding index.
For retrieval, method-level retrieval (i.e., \textit{Mtd}, used for non-chunking prompts) takes 5,468 seconds in total, averaging 2.7 seconds per bug.
Chunk-level retrieval (i.e., \textit{Chk}, used for code-chunking prompts) requires 12,961 seconds in total, averaging 6.5 seconds per bug. The higher cost of chunk-level retrieval is mainly due to the increased number of retrieval queries triggered by fine-grained code chunks.
Note that a bug may require multiple retrievals under chunking. 
Overall, the time overhead introduced by RAG remains moderate, especially considering that retrieval is performed once per bug and does not dominate the overall mutation generation process.


\begin{table}[tp]
  \centering
  \scriptsize
  \caption{GPU Costs of Supervised Fine-Tuning in \ourtool}
    \begin{tabular}{l|r|r|r}
    \hline
    \textbf{Model} & \textbf{Time (s)} & \textbf{FLOPs} & \textbf{GPU Mem.} \\
    \hline
    \textbf{DS-6.7b} & 17,552 & 1.78E+14 & 40GB \\
    \textbf{LM3.1-8b} & 16,267 & 1.48E+14 & 42GB \\
    \textbf{QW2.5-7b} & 14,066 & 1.89E+14 & 40GB \\
    \textbf{QW2.5-14b} & 30,025 & 4.76E+14 & 75GB \\
    \textbf{QW2.5-32b} & 74,493 & 1.05E+15 & 140GB \\
    \hline
    \end{tabular}
  \tablabel{sft-cost}
\end{table}

\tabref{sft-cost} reports the computational cost of supervised fine-tuning (SFT) for different base models.
The SFT time ranges from 14,066 seconds (QW2.5-7b) to 74,493 seconds (QW2.5-32b), with corresponding FLOPs ranging from $1.48\times10^{14}$ to $1.05\times10^{15}$. As expected, both wall-clock time and FLOPs increase substantially with model size.
Peak GPU memory usage varies from 40GB to 42GB for 6-8B models to 75GB for the 14B model and 140GB for the 32B model, indicating that GPU memory becomes the primary hardware constraint for larger models.
Overall, SFT is the most computationally intensive component of \ourtool and requires non-trivial GPU resources, especially for models beyond 14B parameters. However, this cost is incurred only once per model and can be amortized across multiple mutation campaigns.

\subsection{The Influence of RAG Choices}
To examine whether the choice of similarity metric affects \ourtool, we randomly sampled 100 real-world bugs (50 from Defects4J and 50 from ConDefects) and compared three commonly used similarity measures in RAG: cosine similarity, dot-product similarity, and Euclidean distance.

\begin{table}[tp]
  \centering
  \scriptsize
  \caption{Performance of Different RAG Distance Metrics}
    \begin{tabular}{|r|r|r|r|r|r|r|r|r|r|}
    \hline
    \textbf{Method} & \multicolumn{1}{c|}{\textbf{Gen. R.}} & \multicolumn{1}{c|}{\textbf{ND. R.}} & \multicolumn{1}{c|}{\textbf{Com. R.}} & \multicolumn{1}{c|}{\textbf{R. B. D.}} & \multicolumn{1}{c|}{\textbf{O.$\geq$0.8}} & \multicolumn{1}{c|}{\textbf{AOC}} & \multicolumn{1}{c|}{\textbf{APFD-grk}} & \multicolumn{1}{c|}{\textbf{Muse Top-5}} & \multicolumn{1}{c|}{\textbf{Meta Top-5}} \\
    \hline
    \textbf{Cos} & \cellcolor{gray!40}65.7\% & \cellcolor{gray!40}60.2\% & \cellcolor{gray!40}48.5\% & 81 & 34 &  	0.18 &  	0.88 & 51 & 49 \\
    \hline
    \textbf{Dot-product} & 62.0\% & 57.0\% & 46.7\% & \cellcolor{gray!40}85 & 34 & 0.19 & 0.86 & 53 & 52 \\
    \hline
    \textbf{Eucldean} & 63.2\% & 57.2\% & 47.5\% & \cellcolor{gray!40}85 & \cellcolor{gray!40}39 & \cellcolor{gray!40}0.20 & \cellcolor{gray!40}0.9 & \cellcolor{gray!40}56 & \cellcolor{gray!40}54 \\
    \hline
    \end{tabular}
  \tablabel{rag-distance}
\end{table}

As shown in \tabref{rag-distance}, Euclidean distance consistently achieves the best performance across mutation generation quality and downstream applications (MBTCP and MBFL).
Specifically, compared with cosine and dot-product similarity, Euclidean distance improves the coupling rate and real-bug detection rate by approximately 5\%, increases the number of high-Ochiai bugs by 14\%, raises the average Ochiai score by about 10\%, and boosts APFD by 3\%-5\%. It also improves MUSE-Top5 and Meta-Top5 by 9\% and 10\%, respectively. Overall, the results suggest that \ourtool is not overly sensitive to the choice of similarity metric, as all three metrics yield comparable performance trends. Nevertheless, Euclidean distance provides consistently better results and is therefore adopted as the default configuration.

\subsection{Threats to Validity}
Our study faces the following potential threats to validity.

First, the choice of programming language, datasets, mutation generation approaches, and the GPU resource requirements may limit the generalizability of our findings.
We mitigate this by using Java (the most widely studied language), two datasets (Defects4J and ConDefects), and the state-of-the-art LLM-based baselines, along with diverse widely studied LLMs (e.g., GPT-4o, DeepSeek-Coder, Llama-3.1, and Qwen-2.5).

Second, the inherent randomness of LLMs (e.g., decoding strategies) and configuration choices could influence the results. 
To reduce this threat, we follow the default parameters of each model and evaluate multiple open-source LLMs of varying scales as well as GPT-4o.
Another related factor is the number of retrieved few-shot examples used in RAG. Previous work~\cite{wang2024exploratory} reports that varying the number of examples between 5 and 10 does not lead to statistically significant performance differences. Following this finding, we adopt a moderate and commonly used setting within this range. While different retrieval sizes may introduce minor variations, prior evidence suggests that such variations are unlikely to alter the overall conclusions. Nevertheless, we acknowledge that retrieval configuration remains a tunable factor.


Finally, although we carefully controlled the training and evaluation pipeline, potential data leakage remains a general concern in LLM-based studies. To mitigate this threat, we adopted multiple preventive measures.
First, for the RAG retrieval database, we collected data exclusively from real-world GitHub repositories that do not appear in either Defects4J or ConDefects.
We explicitly excluded all projects sharing the same names as those in the two benchmarks to ensure strict dataset separation.
Second, for supervised fine-tuning (SFT), we partitioned projects into disjoint training and testing sets to prevent cross-project leakage.
Specifically, the projects \textit{Lang} and \textit{Cli} were used only to provide training data and were never involved in evaluation.
Third, beyond checking exact matches, we further quantified the similarity between the fine-tuning dataset and the evaluation dataset using N-gram analysis. The results indicate that the two sets are highly dissimilar: the 2-gram similarity is only 1.65\%, and the similarity further drops to 0.019\% for 10-grams, suggesting that long token sequences are rarely shared across datasets. These results provide stronger evidence that memorization-based leakage is unlikely to significantly affect our findings.
Nevertheless, it is still possible that parts of Defects4J or ConDefects may have appeared in the pre-training data of large language models, which is beyond our control.
To estimate its potential impact, we examined whether \ourtool produced exact matches (i.e., syntactically identical mutations) with real bugs.
On Defects4J, the proportions of exact matches remain low across all models, ranging from 1.91\% (GPT-4o) to 2.59\% (DS-6.7b).
On ConDefects, the rates are even lower, all below 0.3\% (e.g., 0.05\% for GPT-4o).
These small percentages suggest that, although data leakage cannot be completely ruled out, its influence on our results is limited.

\section{Related Work} \label{sec:related}

\subsection{Mutation Testing}

Mutation testing is essential for software testing and analysis research, which systematically introduces artificial faults into a program, used to evaluate test effectiveness~\cite{jia2010analysis,papadakis2019mutation}.
Moreover, mutation testing has been widely applied in software engineering, serving as a substitute for real bugs~\cite{just2014mutants,andrews2005mutation,daran1996software,namin2011use}, and supporting tasks such as fault localization~\cite{moon2014ask,papadakis2015metallaxis,wang2025systematic}, test case prioritization~\cite{shin2019empirical}, and automated program repair~\cite{xiao2024accelerating,xiao2023expressapr}.

All downstream applications rely on valid and effective mutations.  
Prior studies suggest that effectiveness increases when mutations are syntactically or semantically close to real bugs~\cite{hindle2016naturalness,jimenez2018mutants,hariri2019comparing}.  
For example, MBFL benefits from mutants that are coupled with real bugs~\cite{moon2014ask,papadakis2015metallaxis,wang2025systematic}.  

\subsection{Mutation Generation} 

Traditional mutation testing relies on \textit{rule-based} operators, i.e., predefined program transformations such as replacing arithmetic or relational operators.  
Most tools (e.g., PIT~\cite{coles2016pit}, Major~\cite{just2011major}, MuJava~\cite{ma2005mujava}, AccMut~\cite{wang2017faster}, WinMut~\cite{wang2021faster}, Milu~\cite{jia2008milu}) follow this paradigm.  

With the rise of machine learning, \textit{learning-based} methods emerged.  
Examples include DeepMutation~\cite{tufano2019learning} using seq2seq translation, LEAM~\cite{tian2022learning} with learning-guided rule expansion, and $\mu$BERT~\cite{degiovanni2022mubert}, which predicts masked code tokens via BERT~\cite{devlin2018bert}.  
However, such approaches often struggle to ensure syntactic correctness and mutation diversity.  

Recently, \textit{LLM-based} methods have attracted growing attention~\cite{tip2025llmorpheus,deb2024syntax,endres2024can,dakhel2024effective,ibrahimzada2023automated,wang2024exploratory}.  
These studies show the feasibility of leveraging powerful generative models for mutation generation, though most focus on limited metrics such as mutation score or repair effectiveness.  
In contrast, our work systematically evaluates diverse LLMs and prompt strategies against rule-based and learning-based baselines, measuring both the validity and effectiveness of the generated mutants in multiple downstream tasks.

\section{Conclusion} \label{sec:conclusion}

In this paper, we propose \ourtool, a novel LLM-based mutation generation framework that combines retrieval-augmented generation, code chunking, and supervised fine-tuning.  
Through extensive experiments on 1,991 real-world Java bugs from Defects4J and ConDefects, we demonstrated that \ourtool consistently generates more valid and effective mutations than state-of-the-art baselines.  
It significantly improves mutation validity (e.g., generation rate from 42.89\% to 65.6\%) and effectiveness (e.g., real bug detection rate of 92.61\% vs.\ 57.86\% for LLMut), while also boosting the performance of downstream applications such as test case prioritization and fault localization.  
Moreover, our results show that \ourtool enables 7B-scale open-source models to achieve performance competitive with, and in some cases surpassing, GPT-4o, highlighting its strong practical value.



\bibliographystyle{ACM-Reference-Format}
\bibliography{ref}

@inproceedings{harman2025mutation,
  title={Mutation-Guided LLM-based Test Generation at Meta},
  author={Harman, Mark and Ritchey, Jillian and Harper, Inna and Sengupta, Shubho and Mao, Ke and Gulati, Abhishek and Foster, Christopher and Robert, Herv\'{e}},
  booktitle={Proceedings of the 33rd ACM International Conference on the Foundations of Software Engineering},
  pages={180--191},
  year={2025}
}

@article{endres2024can,
  title={Can Large Language Models Transform Natural Language Intent into Formal Method Postconditions?},
  author={Endres, Madeline and Fakhoury, Sarah and Chakraborty, Saikat and Lahiri, Shuvendu K},
  journal={Proceedings of the ACM on Software Engineering},
  volume={1},
  number={FSE},
  pages={1889--1912},
  year={2024},
  publisher={ACM New York, NY, USA}
}

@article{chen2024deep,
  title={Deep learning-based software engineering: progress, challenges, and opportunities},
  author={Chen, Xiangping and Hu, Xing and Huang, Yuan and Jiang, He and Ji, Weixing and Jiang, Yanjie and Jiang, Yanyan and Liu, Bo and Liu, Hui and Li, Xiaochen and others},
  journal={Science China Information Sciences},
  volume={68},
  number={1},
  pages={111102},
  year={2025},
  publisher={Springer}
}

@article{demillo1978hints,
  title={Hints on test data selection: Help for the practicing programmer},
  author={DeMillo, Richard A and Lipton, Richard J and Sayward, Frederick G},
  journal={Computer},
  volume={11},
  number={4},
  pages={34--41},
  year={1978},
  publisher={IEEE}
}

@article{zhu2024deepseek,
  title={DeepSeek-Coder-V2: Breaking the Barrier of Closed-Source Models in Code Intelligence},
  author={Zhu, Qihao and Guo, Daya and Shao, Zhihong and Yang, Dejian and Wang, Peiyi and Xu, Runxin and Wu, Y and Li, Yukun and Gao, Huazuo and Ma, Shirong and others},
  journal={arXiv preprint arXiv:2406.11931},
  year={2024}
}

@article{hamlet1977testing,
  title={Testing programs with the aid of a compiler},
  author={Hamlet, Richard G.},
  journal={IEEE transactions on software engineering},
  number={4},
  pages={279--290},
  year={1977},
  publisher={IEEE}
}

@article{jia2010analysis,
  title={An analysis and survey of the development of mutation testing},
  author={Jia, Yue and Harman, Mark},
  journal={IEEE transactions on software engineering},
  volume={37},
  number={5},
  pages={649--678},
  year={2010},
  publisher={IEEE}
}

@incollection{papadakis2019mutation,
	title={Mutation testing advances: an analysis and survey},
	author={Papadakis, Mike and Kintis, Marinos and Zhang, Jie and Jia, Yue and Le Traon, Yves and Harman, Mark},
	booktitle={Advances in Computers},
	volume={112},
	pages={275--378},
	year={2019},
	publisher={Elsevier}
}

@inproceedings{just2014mutants,
  title={Are mutants a valid substitute for real faults in software testing?},
  author={Just, Ren{\'e} and Jalali, Darioush and Inozemtseva, Laura and Ernst, Michael D and Holmes, Reid and Fraser, Gordon},
  booktitle={Proceedings of the 22nd ACM SIGSOFT International Symposium on Foundations of Software Engineering},
  pages={654--665},
  year={2014}
}

@inproceedings{coles2016pit,
  title={Pit: a practical mutation testing tool for java},
  author={Coles, Henry and Laurent, Thomas and Henard, Christopher and Papadakis, Mike and Ventresque, Anthony},
  booktitle={Proceedings of the 25th international symposium on software testing and analysis},
  pages={449--452},
  year={2016}
}

@inproceedings{just2011major,
  title={MAJOR: An efficient and extensible tool for mutation analysis in a {Java} compiler},
  author={Just, Rene and Schweiggert, Franz and Kapfhammer, Gregory M},
  booktitle={ASE},
  pages={612--615},
  year={2011},
}

@inproceedings{tian2022learning,
  title={Learning to construct better mutation faults},
  author={Tian, Zhao and Chen, Junjie and Zhu, Qihao and Yang, Junjie and Zhang, Lingming},
  booktitle={Proceedings of the 37th IEEE/ACM International Conference on Automated Software Engineering},
  pages={1--13},
  year={2022}
}

@inproceedings{tufano2019learning,
  title={Learning how to mutate source code from bug-fixes},
  author={Tufano, Michele and Watson, Cody and Bavota, Gabriele and Di Penta, Massimiliano and White, Martin and Poshyvanyk, Denys},
  booktitle={2019 IEEE International conference on software maintenance and evolution (ICSME)},
  pages={301--312},
  year={2019},
  organization={IEEE}
}

@inproceedings{wang2017faster,
  title={Faster mutation analysis via equivalence modulo states},
  author={Wang, Bo and Xiong, Yingfei and Shi, Yangqingwei and Zhang, Lu and Hao, Dan},
  booktitle={Proceedings of the 26th ACM SIGSOFT International Symposium on Software Testing and Analysis},
  pages={295--306},
  year={2017}
}

@inproceedings{wang2021faster,
  title={Faster mutation analysis with fewer processes and smaller overheads},
  author={Wang, Bo and Lu, Sirui and Xiong, Yingfei and Liu, Feng},
  booktitle={2021 36th IEEE/ACM International Conference on Automated Software Engineering (ASE)},
  pages={381--393},
  year={2021},
  organization={IEEE}
}

@article{budd1982two,
  title={Two notions of correctness and their relation to testing},
  author={Budd, Timothy A and Angluin, Dana},
  journal={Acta informatica},
  volume={18},
  number={1},
  pages={31--45},
  year={1982},
  publisher={Springer}
}

@article{achiam2023gpt,
  title={Gpt-4 technical report},
  author={Achiam, Josh and Adler, Steven and Agarwal, Sandhini and Ahmad, Lama and Akkaya, Ilge and Aleman, Florencia Leoni and Almeida, Diogo and Altenschmidt, Janko and Altman, Sam and Anadkat, Shyamal and others},
  journal={arXiv preprint arXiv:2303.08774},
  year={2023}
}

@article{roziere2023code,
  title={Code llama: Open foundation models for code},
  author={Roziere, Baptiste and Gehring, Jonas and Gloeckle, Fabian and Sootla, Sten and Gat, Itai and Tan, Xiaoqing Ellen and Adi, Yossi and Liu, Jingyu and Remez, Tal and Rapin, J{\'e}r{\'e}my and others},
  journal={arXiv preprint arXiv:2308.12950},
  year={2023}
}

@inproceedings{just2014defects4j,
  title={Defects4J: A database of existing faults to enable controlled testing studies for Java programs},
  author={Just, Ren{\'e} and Jalali, Darioush and Ernst, Michael D},
  booktitle={Proceedings of the 2014 International Symposium on Software Testing and Analysis},
  pages={437--440},
  year={2014}
}

@inproceedings{wu2023condefects,
  title={ConDefects: A Complementary Dataset to Address the Data Leakage Concern for LLM-Based Fault Localization and Program Repair},
  author={Wu, Yonghao and Li, Zheng and Zhang, Jie M and Liu, Yong},
  booktitle={Companion Proceedings of the 32nd ACM International Conference on the Foundations of Software Engineering},
  pages={642--646},
  year={2024}
}

@article{hindle2016naturalness,
  title={On the naturalness of software},
  author={Hindle, Abram and Barr, Earl T and Gabel, Mark and Su, Zhendong and Devanbu, Premkumar},
  journal={Communications of the ACM},
  volume={59},
  number={5},
  pages={122--131},
  year={2016},
  publisher={ACM New York, NY, USA}
}

@article{ojdanic2023syntactic,
  title={Syntactic vs. semantic similarity of artificial and real faults in mutation testing studies},
  author={Ojdanic, Milos and Garg, Aayush and Khanfir, Ahmed and Degiovanni, Renzo and Papadakis, Mike and Le Traon, Yves},
  journal={IEEE Transactions on Software Engineering},
  year={2023},
  publisher={IEEE}
}

@inproceedings{jimenez2018mutants,
  title={Are mutants really natural? a study on how" naturalness" helps mutant selection},
  author={Jimenez, Matthieu and Checkam, Thiery Titcheu and Cordy, Maxime and Papadakis, Mike and Kintis, Marinos and Traon, Yves Le and Harman, Mark},
  booktitle={Proceedings of the 12th ACM/IEEE International Symposium on Empirical Software Engineering and Measurement},
  pages={1--10},
  year={2018}
}

@inproceedings{kaufman2022prioritizing,
  title={Prioritizing mutants to guide mutation testing},
  author={Kaufman, Samuel J and Featherman, Ryan and Alvin, Justin and Kurtz, Bob and Ammann, Paul and Just, Ren{\'e}},
  booktitle={Proceedings of the 44th International Conference on Software Engineering},
  pages={1743--1754},
  year={2022}
}

@inproceedings{degiovanni2022mubert,
  title={$\mu$bert: Mutation testing using pre-trained language models},
  author={Degiovanni, Renzo and Papadakis, Mike},
  booktitle={2022 IEEE International Conference on Software Testing, Verification and Validation Workshops (ICSTW)},
  pages={160--169},
  year={2022},
  organization={IEEE}
}

@article{kim2022predictive,
  title={Predictive mutation analysis via the natural language channel in source code},
  author={Kim, Jinhan and Jeon, Juyoung and Hong, Shin and Yoo, Shin},
  journal={ACM Transactions on Software Engineering and Methodology (TOSEM)},
  volume={31},
  number={4},
  pages={1--27},
  year={2022},
  publisher={ACM New York, NY}
}

@article{devlin2018bert,
  title={Bert: Pre-training of deep bidirectional transformers for language understanding},
  author={Devlin, Jacob and Chang, Ming-Wei and Lee, Kenton and Toutanova, Kristina},
  journal={arXiv preprint arXiv:1810.04805},
  year={2018}
}

@inproceedings{andrews2005mutation,
  title={Is mutation an appropriate tool for testing experiments?},
  author={Andrews, James H and Briand, Lionel C and Labiche, Yvan},
  booktitle={Proceedings of the 27th international conference on Software engineering},
  pages={402--411},
  year={2005}
}

@article{daran1996software,
  title={Software error analysis: A real case study involving real faults and mutations},
  author={Daran, Murial and Th{\'e}venod-Fosse, Pascale},
  journal={ACM SIGSOFT Software Engineering Notes},
  volume={21},
  number={3},
  pages={158--171},
  year={1996},
  publisher={ACM New York, NY, USA}
}

@inproceedings{namin2011use,
  title={The use of mutation in testing experiments and its sensitivity to external threats},
  author={Namin, Akbar Siami and Kakarla, Sahitya},
  booktitle={Proceedings of the 2011 International Symposium on Software Testing and Analysis},
  pages={342--352},
  year={2011}
}

@article{xiao2024accelerating,
  title={Accelerating patch validation for program repair with interception-based execution scheduling},
  author={Xiao, Yuan-An and Yang, Chenyang and Wang, Bo and Xiong, Yingfei},
  journal={IEEE Transactions on Software Engineering},
  year={2024},
  publisher={IEEE}
}

@inproceedings{xiao2023expressapr,
  title={ExpressAPR: Efficient patch validation for Java automated program repair systems},
  author={Xiao, Yuan-An and Yang, Chenyang and Wang, Bo and Xiong, Yingfei},
  booktitle={2023 38th IEEE/ACM International Conference on Automated Software Engineering (ASE)},
  pages={2038--2041},
  year={2023},
  organization={IEEE}
}

@inproceedings{moon2014ask,
  title={Ask the mutants: Mutating faulty programs for fault localization},
  author={Moon, Seokhyeon and Kim, Yunho and Kim, Moonzoo and Yoo, Shin},
  booktitle={2014 IEEE Seventh International Conference on Software Testing, Verification and Validation},
  pages={153--162},
  year={2014},
  organization={IEEE}
}

@article{papadakis2015metallaxis,
  title={Metallaxis-FL: mutation-based fault localization},
  author={Papadakis, Mike and Le Traon, Yves},
  journal={Software Testing, Verification and Reliability},
  volume={25},
  number={5-7},
  pages={605--628},
  year={2015},
  publisher={Wiley Online Library}
}

@article{shin2019empirical,
  title={Empirical evaluation of mutation-based test case prioritization techniques},
  author={Shin, Donghwan and Yoo, Shin and Papadakis, Mike and Bae, Doo-Hwan},
  journal={Software Testing, Verification and Reliability},
  volume={29},
  number={1-2},
  pages={e1695},
  year={2019},
  publisher={Wiley Online Library}
}

@inproceedings{beller2021would,
  title={What it would take to use mutation testing in industry—a study at facebook},
  author={Beller, Moritz and Wong, Chu-Pan and Bader, Johannes and Scott, Andrew and Machalica, Mateusz and Chandra, Satish and Meijer, Erik},
  booktitle={2021 IEEE/ACM 43rd International Conference on Software Engineering: Software Engineering in Practice (ICSE-SEIP)},
  pages={268--277},
  year={2021},
  organization={IEEE}
}

@article{petrovic2021practical,
  title={Practical mutation testing at scale: A view from google},
  author={Petrovi{\'c}, Goran and Ivankovi{\'c}, Marko and Fraser, Gordon and Just, Ren{\'e}},
  journal={IEEE Transactions on Software Engineering},
  volume={48},
  number={10},
  pages={3900--3912},
  year={2021},
  publisher={IEEE}
}

@inproceedings{jia2008milu,
  title={MILU: A customizable, runtime-optimized higher order mutation testing tool for the full C language},
  author={Jia, Yue and Harman, Mark},
  booktitle={Testing: Academic \& Industrial Conference-Practice and Research Techniques (taic part 2008)},
  pages={94--98},
  year={2008},
  organization={IEEE}
}

@article{wang2024software,
  title={Software testing with large language models: Survey, landscape, and vision},
  author={Wang, Junjie and Huang, Yuchao and Chen, Chunyang and Liu, Zhe and Wang, Song and Wang, Qing},
  journal={IEEE Transactions on Software Engineering},
  year={2024},
  publisher={IEEE}
}

@article{deb2024syntax,
  title={Syntax is all you need: A universal-language approach to mutant generation},
  author={Deb, Sourav and Jain, Kush and Van Tonder, Rijnard and Le Goues, Claire and Groce, Alex},
  journal={Proceedings of the ACM on Software Engineering},
  volume={1},
  number={FSE},
  pages={654--674},
  year={2024},
  publisher={ACM New York, NY, USA}
}

@inproceedings{deng2024large,
  title={Large language models are edge-case generators: Crafting unusual programs for fuzzing deep learning libraries},
  author={Deng, Yinlin and Xia, Chunqiu Steven and Yang, Chenyuan and Zhang, Shizhuo Dylan and Yang, Shujing and Zhang, Lingming},
  booktitle={Proceedings of the 46th IEEE/ACM International Conference on Software Engineering},
  pages={1--13},
  year={2024}
}

@article{chen2021evaluating,
  title={Evaluating large language models trained on code},
  author={Chen, Mark and Tworek, Jerry and Jun, Heewoo and Yuan, Qiming and Pinto, Henrique Ponde de Oliveira and Kaplan, Jared and Edwards, Harri and Burda, Yuri and Joseph, Nicholas and Brockman, Greg and others},
  journal={arXiv preprint arXiv:2107.03374},
  year={2021}
}

@inproceedings{hariri2019comparing,
  title={Comparing mutation testing at the levels of source code and compiler intermediate representation},
  author={Hariri, Farah and Shi, August and Fernando, Vimuth and Mahmood, Suleman and Marinov, Darko},
  booktitle={2019 12th IEEE conference on software testing, validation and verification (ICST)},
  pages={114--124},
  year={2019},
  organization={IEEE}
}

@article{ma2005mujava,
  title={MuJava: an automated class mutation system},
  author={Ma, Yu-Seung and Offutt, Jeff and Kwon, Yong Rae},
  journal={Software Testing, Verification and Reliability},
  volume={15},
  number={2},
  pages={97--133},
  year={2005},
  publisher={Wiley Online Library}
}

@inproceedings{tian2024large,
  title={Large Language Models for Equivalent Mutant Detection: How Far Are We?},
  author={Tian, Zhao and Shu, Honglin and Wang, Dong and Cao, Xuejie and Kamei, Yasutaka and Chen, Junjie},
  booktitle={Proceedings of the 33rd ACM SIGSOFT International Symposium on Software Testing and Analysis},
  pages={1733--1745},
  year={2024}
}

@article{zou2019empirical,
  title={An empirical study of fault localization families and their combinations},
  author={Zou, Daming and Liang, Jingjing and Xiong, Yingfei and Ernst, Michael D and Zhang, Lu},
  journal={IEEE Transactions on Software Engineering},
  volume={47},
  number={2},
  pages={332--347},
  year={2019},
  publisher={IEEE}
}

@article{wang2025systematic,
  title={A Systematic Exploration of Mutation-Based Fault Localization Formulae},
  author={Wang, Bo and Wei, Jinkang and Chen, Mingda and Chen, Chong and Lin, Youfang and Zhang, Jie M},
  journal={Software Testing, Verification and Reliability},
  volume={35},
  number={1},
  pages={e1905},
  year={2025},
  publisher={Wiley Online Library}
}

@article{dakhel2024effective,
  title={Effective test generation using pre-trained large language models and mutation testing},
  author={Dakhel, Arghavan Moradi and Nikanjam, Amin and Majdinasab, Vahid and Khomh, Foutse and Desmarais, Michel C},
  journal={Information and Software Technology},
  volume={171},
  pages={107468},
  year={2024},
  publisher={Elsevier}
}

@article{wang2024exploratory,
  title={A Comprehensive Study on Large Language Models for Mutation Testing},
  author={Wang, Bo and Chen, Mingda and Deng, Ming and Lin, Youfang and Harman, Mark and Papadakis, Mike and Zhang, Jie M},
  journal={ACM Transactions on Software Engineering and Methodology (TOSEM)},
  year = {2026},
  publisher = {Association for Computing Machinery},
  address = {New York, NY, USA}
}

@article{ibrahimzada2023automated,
  title={Automated bug generation in the era of large language models},
  author={Ibrahimzada, Ali Reza and Chen, Yang and Rong, Ryan and Jabbarvand, Reyhaneh},
  journal={arXiv preprint arXiv:2310.02407},
  year={2023}
}

@article{tianleam++,
  title={LEAM++: Learning for Selective Mutation Fault Construction},
  author={Tian, Zhao and Chen, Junjie and Wang, Dong and Zhu, Qihao and Fan, Xingyu and Zhang, Lingming},
  journal={ACM Transactions on Software Engineering and Methodology},
  publisher={ACM New York, NY},
  year = {2025},
  volume = {34},
  number = {8},
  numpages = {41},
  articleno = {223},
}

@article{tip2025llmorpheus,
  title={Llmorpheus: Mutation testing using large language models},
  author={Tip, Frank and Bell, Jonathan and Sch{\"a}fer, Max},
  journal={IEEE Transactions on Software Engineering},
  year={2025},
  publisher={IEEE}
}

@article{do2006use,
  title={On the use of mutation faults in empirical assessments of test case prioritization techniques},
  author={Do, Hyunsook and Rothermel, Gregg},
  journal={IEEE Transactions on Software Engineering},
  volume={32},
  number={9},
  pages={733--752},
  year={2006},
  publisher={IEEE}
}

@inproceedings{lou2015mutation,
  title={Mutation-based test-case prioritization in software evolution},
  author={Lou, Yiling and Hao, Dan and Zhang, Lu},
  booktitle={2015 IEEE 26th International Symposium on Software Reliability Engineering (ISSRE)},
  pages={46--57},
  year={2015},
  organization={IEEE}
}

@article{lewis2020retrieval,
  title={Retrieval-augmented generation for knowledge-intensive nlp tasks},
  author={Lewis, Patrick and Perez, Ethan and Piktus, Aleksandra and Petroni, Fabio and Karpukhin, Vladimir and Goyal, Naman and K{\"u}ttler, Heinrich and Lewis, Mike and Yih, Wen-tau and Rockt{\"a}schel, Tim and others},
  journal={Advances in neural information processing systems},
  volume={33},
  pages={9459--9474},
  year={2020}
}

@inproceedings{wang2024hits,
  title={Hits: High-coverage llm-based unit test generation via method slicing},
  author={Wang, Zejun and Liu, Kaibo and Li, Ge and Jin, Zhi},
  booktitle={Proceedings of the 39th IEEE/ACM International Conference on Automated Software Engineering},
  pages={1258--1268},
  year={2024}
}

@article{wang2023codet5,
  title={Codet5+: Open code large language models for code understanding and generation},
  author={Wang, Yue and Le, Hung and Gotmare, Akhilesh Deepak and Bui, Nghi DQ and Li, Junnan and Hoi, Steven CH},
  journal={arXiv preprint arXiv:2305.07922},
  year={2023}
}

@article{wang2021codet5,
  title={Codet5: Identifier-aware unified pre-trained encoder-decoder models for code understanding and generation},
  author={Wang, Yue and Wang, Weishi and Joty, Shafiq and Hoi, Steven CH},
  journal={arXiv preprint arXiv:2109.00859},
  year={2021}
}

@article{yoo2012regression,
  title={Regression testing minimization, selection and prioritization: a survey},
  author={Yoo, Shin and Harman, Mark},
  journal={Software testing, verification and reliability},
  volume={22},
  number={2},
  pages={67--120},
  year={2012},
  publisher={Wiley Online Library}
}

@article{rothermel2001prioritizing,
  title={Prioritizing test cases for regression testing},
  author={Rothermel, Gregg and Untch, Roland H. and Chu, Chengyun and Harrold, Mary Jean},
  journal={IEEE Transactions on software engineering},
  volume={27},
  number={10},
  pages={929--948},
  year={2001},
  publisher={IEEE}
}

@inproceedings{rothermel1999test,
  title={Test case prioritization: An empirical study},
  author={Rothermel, Gregg and Untch, Roland H and Chu, Chengyun and Harrold, Mary Jean},
  booktitle={Proceedings IEEE International Conference on Software Maintenance-1999 (ICSM'99).'Software Maintenance for Business Change'(Cat. No. 99CB36360)},
  pages={179--188},
  year={1999},
  organization={IEEE}
}

@article{yang2025qwen3,
  title={Qwen3 technical report},
  author={Yang, An and Li, Anfeng and Yang, Baosong and Zhang, Beichen and Hui, Binyuan and Zheng, Bo and Yu, Bowen and Gao, Chang and Huang, Chengen and Lv, Chenxu and others},
  journal={arXiv preprint arXiv:2505.09388},
  year={2025}
}

@article{grattafiori2024llama,
  title={The llama 3 herd of models},
  author={Grattafiori, Aaron and Dubey, Abhimanyu and Jauhri, Abhinav and Pandey, Abhinav and Kadian, Abhishek and Al-Dahle, Ahmad and Letman, Aiesha and Mathur, Akhil and Schelten, Alan and Vaughan, Alex and others},
  journal={arXiv preprint arXiv:2407.21783},
  year={2024}
}

@inproceedings{chen2025camus,
  title={CAMUS: Context-Aware Neural Mutation Selection},
  author={Chen, Mingda and Wang, Bo and Lin, Youfang and Zhang, Jie M},
  booktitle={2025 32nd Asia-Pacific Software Engineering Conference (APSEC)},
  pages={11--22},
  year={2025},
  organization={IEEE}
}

\end{document}